\documentclass[sigconf]{acmart}

\setcopyright{none}
\renewcommand\footnotetextcopyrightpermission[1]{}

\copyrightyear{2025} 
\acmYear{2025} 
\acmConference[]{Preprint}{2025}{ }
\acmBooktitle{ACM eEnergy}

\usepackage{graphicx}  
\usepackage{algorithm,algpseudocode} 
\usepackage{algorithmicx}

\usepackage{etex}
\usepackage{enumitem}
\usepackage{booktabs}
\usepackage{amsmath,amsfonts,mathtools,mathrsfs}
\usepackage{siunitx}

\usepackage[sans]{dsfont}
\usepackage{stmaryrd}
\usepackage{pifont}
\usepackage{wasysym}
\usepackage{colortbl}
\usepackage{array}
	
\usepackage{amstext,xspace}
\usepackage{multirow}
\usepackage{acmart-taps}
\newcommand*\rot[1]{\rotatebox{30}{\makebox[0pt][l]{#1}}}

\newcommand{\xmark}{\ding{55}}

\usepackage{multirow}
\usepackage{multicol}
\usepackage{caption}
\usepackage{tabularx}
\usepackage{subfigure}

\usepackage{tikz}
\usepackage{forest}
\usetikzlibrary{trees,positioning,shapes,shadows,arrows.meta}

\begin{document}

\title{Destabilizing Power Grid and Energy Market by Cyberattacks on Smart Inverters}

\author{Xiangyu Hui}
\orcid{0009-0007-8054-1841}
\affiliation{\institution{University of Melbourne}
\city{Melbourne}
\country{Australia}}
\email{xiangyu.hui.1@}
\email{student.unimelb.edu.au}

\author{Samuel Karumba}
\orcid{0000-0002-7181-9799}
\affiliation{\institution{CSIRO Data61}
\city{Sydney}
\country{Australia}}
%\email{samuel.karumba@data61.csiro.au}

\author{Sid Chi-Kin Chau}\authornote{Corresponding author: \url{sid.chau@acm.org}}
\orcid{0000-0003-0362-2844}
\affiliation{\institution{CSIRO Data61}
\city{Sydney}
\country{Australia}}
\email{sid.chau@acm.org}

\author{Mohiuddin Ahmed}
\orcid{0000-0003-2028-6475}
\affiliation{\institution{Edith Cowan University}
\city{Perth}
\country{Australia}}
\email{mohiuddin.ahmed@ecu.edu.au}

\pagestyle{plain}
\thispagestyle{plain}

\begin{abstract}
Cyberattacks on smart inverters and distributed PV are becoming an imminent threat, because of the recent well-documented vulnerabilities and attack incidents. Particularly, the long lifespan of inverter devices, users' oblivion of cybersecurity compliance, and the lack of cyber regulatory frameworks exacerbate the prospect of cyberattacks on smart inverters. As a result, this raises a question -- {\em do cyberattacks on smart inverters, if orchestrated on a large scale, pose a genuine threat of wide-scale instability to the power grid and energy market}? This paper provides a realistic assessment on the plausibility and impacts of wide-scale power instability caused by cyberattacks on smart inverters. We conduct an in-depth study based on the electricity market data of Australia and the knowledge of practical contingency mechanisms. Our key findings reveal: (1) Despite the possibility of disruption to the grid by cyberattacks on smart inverters, the impact is only significant under careful planning and orchestration. (2) While the grid can assure certain power system security to survive inadvertent contingency events, it is insufficient to defend against savvy attackers who can orchestrate attacks in an adversarial manner. Our data analysis of Australia's electricity grid also reveals that a relatively low percentage of distributed PV would be sufficient to launch an impactful concerted attack on the grid. Our study casts insights on robust strategies for defending the grid in the presence of cyberattacks for places with high penetration of distributed PV.\footnote{This is an extended version of the conference paper \cite{HKCA25inverter} appearing in ACM International Conference on Future and Sustainable Energy Systems (ACM e-Energy 2025).}\enlargethispage{15pt}
\end{abstract}

\keywords{Cyberattacks on Power Grid, Energy Cybersecurity, Smart Inverters}

\maketitle

%\vspace{-5pt}
\section{Introduction} \label{sec:intro}

The rapid rise of distributed photovoltaic (DPV) systems, equipped with smart inverters for intelligent power conversion and advanced grid-supporting functions, is transforming the energy infrastructure into an open, decentralized architecture. These {\em inverter-based consumer energy resources} are expected to contribute a substantial portion of energy to the power grid and will become the primary energy source around the world in the near future \cite{nijsse2023momentum}. For example, Australia is projected to reach 45\% of all electricity to be generated from consumer energy resources in the coming decades, dominated by rooftop solar from households and small businesses \cite{CSIRO_ENA_2017}. \enlargethispage{15pt}

The abundance of inverter-based consumer energy resources will pose unprecedented threats to the reliability of the grid, not only because of the nature of intermittent energy generation but also due to the decentralized, often unregulated control by diverse participants, ranging from users to vendors and manufacturers. These devices are usually developed and manufactured overseas and critically rely on operational support and software maintenance provided by overseas manufacturers. Smart inverters need to interact with diverse systems and parties, inducing new attack vectors to be exploited by malicious actors to disrupt the energy infrastructure. Particularly, the long lifespan of typical consumer energy devices in operation, users' oblivion on cybersecurity compliance, and the lack of related cyber regulatory frameworks will aggravate the likelihood of cyberattacks on smart inverters \cite{KCPAJ24inverter}. 

Therefore, this raises a crucial question -- {\bf\em do cyberattacks on smart inverters, even if they can be orchestrated on a large scale, pose a genuine threat of wide-scale instability to the power grid and energy market}\footnote{Note that the wide-scale power outage in Spain and Portugal on 28 April 2025 highlights the severe impacts that power grid instability can cause to the society \cite{SpainPortOutage}.}? 
\begin{itemize}

\item[$\triangleright${\bf\em Yes}:]
On one hand, the prospect of cyberattacks on a significant number of smart inverts is increasingly plausible, because of the well-documented vulnerabilities and incidents of attacks on smart inverts. Also, there is a likelihood of botched software patching (e.g., CrowdStrike) and wide DDoS attacks by botnets (e.g., Mirai) that can bring large-scale failures to distributed energy devices. Given the significant reliance of energy from inverter-based resources to the grid, one argues that the impact of  concerted cyberattacks on smart inverters would be severe and possibly lead to a collapse of major energy supplies, causing large-scale power outages. 

\item[$\triangleright${\bf\em No}:]
On the other hand, the likelihood of large-scale power outages caused by smart inverter failures is disputed from the power system security perspective. Modern power grids are already designed to withstand contingency events like a power plant failure or a shortfall of solar energy due to weather conditions. The energy supplies are typically well-provisioned by contingency mechanisms (e.g., by fossil-fuelled generation), such that the grid should be able to provide sufficient energy in the absence of solar energy, particularly during night-time. Hence, cyber threats to smart inverters would not significantly disrupt power stability. 

\end{itemize} 

In this paper, we give a critical assessment of the plausibility and impacts of wide-scale power instability caused by cyberattacks on smart inverters. Our answer to the above question is ``{\em yes-and-no}''. Our key findings reveal: (1) Despite the possibility of disruption to the grid by cyberattacks on smart inverters, the impact is only significant under careful planning and orchestration. (2) While the grid can assure certain power system security to survive inadvertent contingency events, it is not sufficient to defend against savvy attackers who can orchestrate attacks in an adversarial manner. In contrast with the prior assessment studies based on simplified assumptions and back-of-the-envelope calculations \cite{musleh2024experimental, goerke2024controls}, we conduct an in-depth study based on the electricity market data of Australia and the knowledge of practical contingency mechanisms. Our data analysis of Australia's electricity grid also reveals that a relatively low percentage of DPV would be sufficient to launch an impactful concerted attack on the grid. Therefore, our study casts insights on robust strategies for defending the grid in the presence of possible concerted cyberattacks on smart inverters and DPV.

Our study, though based on Australia's power grid and electricity market, can provide general insights for several reasons: (1) Australia is a pioneer for market-based co-optimized frequency control. While there are other places that also adopt co-optimized frequency control \cite{lal2021essential} (e.g., by California ISO (CAISO), Texas ISO (ERCOT), UK ISO and Ireland ISO), there is a lack of transparency regarding contingency generation in other places. Our study provides insights into how co-optimized frequency control may be susceptible to cyberattacks, which sheds light on the robust adoption of co-optimized frequency control in other places. (2) Australia has a high penetration of DPV (with 20GW+ installed solar capacity out of 65GW total generation capacity), supplying up to 75\% of total energy in several states \cite{australia2017peak}, which is ahead of many countries. The lessons learnt from Australia will be instrumental to many DPV-populated places in the world. 

\smallskip\enlargethispage{15pt}
\noindent
$\triangleright$ {\bf  Summary of Findings:}
We summarize the general findings based on our assessment of Australia's electricity grid as follows: 
\begin{enumerate}[leftmargin=*]

\item {\bf\em Misalignment of Contingency Capacity:} The existing mechanisms for contingency generation and frequency control typically consider inadvertent events like failure of a fossil-fueled generator. It does not align with the DPV generation. Hence, attackers can exploit the lack of contingency capacity during the high levels of DPV generation to launch a cyberattack with only a small portion of DPV.

\item {\bf\em Impacts of Possible Attacks:} If savvy attackers orchestrate attacks by considering the levels of contingency capacity and the amount of synchronous inertia (which correlates with fossil-fueled generation), this will create significant impacts on the grid by destabilizing the system frequency in rapid timescale.

\item {\bf\em Predictability of Opportunities:} Attackers can further maximize the impacts by leveraging machine learning to predict the optimal opportunities from open market data and facility data (e.g., planned outages), which will be hard to defend by the existing contingency mechanisms.

\end{enumerate}

Note that there are limitations in our findings. See a detailed discussion in Sec.~\ref{sec:discuss}. But our study identifies the fundamental impacts and ramifications of plausible cyberattacks on smart inverters.

\smallskip
\noindent
$\triangleright$ {\bf  Paper Organization:} Sec.~\ref{sec:bg} presents the background of cyberattacks on smart inverters and power system security, as well as a brief survey of the related work. Sec.~\ref{sec:wem} presents an overview of Australia’s electricity market and its contingency mechanism. Sec.~\ref{sec:analysis} provides a detailed assessment of the plausibility and impacts of wide-scale instability to Australia’s grid caused by possible concerted cyberattacks on smart inverters. Sec.~\ref{sec:discuss} provides a discussion on the limitations, mitigation strategies and implications.

\enlargethispage{15pt}
\section{Background and Related Work} \label{sec:bg}

\subsection{Overview of Smart Inverters}

In this section, we provide an overview of smart inverters. Smart inverters substantially expand the basic function of traditional inverters of DC-AC electricity conversion with advanced capabilities of supporting the power grid, such as real-time solar energy monitoring, remote device control, low-voltage ride-through, generation regulation and demand response management. A smart inverter consists of a software-defined control unit that allows its control logic (e.g., operational parameters, constraints) to be modifiable through firmware upgrades and a wireless/wireline communication module that allows interconnections with users, grid operators, manufacturers, aggregators and other energy management systems.

\begin{figure}[t!]
\centering  \vspace{-5pt}
\includegraphics[scale=1.3]{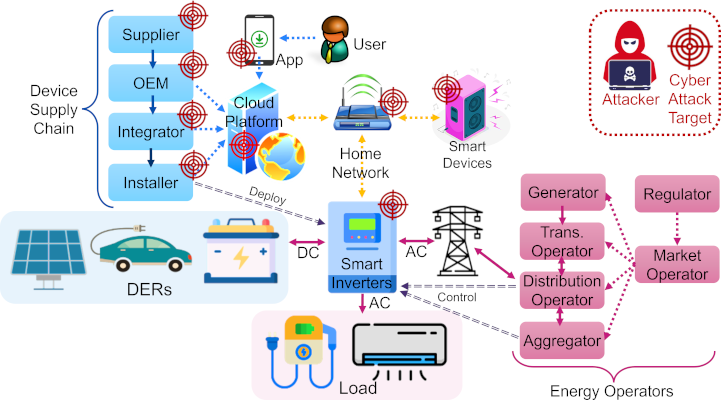}  \vspace{-10pt}
\caption{An illustration of the systems and parties interacting with a smart inverter.} 
\label{fig:system}  \vspace{-15pt}
\end{figure}

Smart inverters interact with diverse systems and parties, such as (1) {\em home energy systems} including distributed energy resources (DERs) and appliance load, (2) {\em energy operators} for the delivery and control of energy from generation to distribution, (3) {\em device supply chain} for the manufacturing, installation and maintenance of inverter devices, most of which can access smart inverters through cloud platforms during their lifetime, (4) {\em home smart devices} that share the same home networks with smart inverters. See Fig~\ref{fig:system} for an illustration of the systems and parties interacting with a smart inverter. Note that attack opportunities can be exploited by targeting those interconnected systems to disrupt smart inverters and subsequently the energy infrastructure \cite{KCPAJ24inverter}. In the following, we survey several reported vulnerabilities of smart inverters.

\vspace{-5pt}
\subsection{Reported Vulnerabilities of Smart Inverters}

Since a smart inverter is also a smart connected home device, it can be exploited in a similar manner as many smart home devices. For instance, attackers can subvert the control unit through malicious or botched system updates, manipulate the communication between the device and a remote party by replay attacks using intercepted data, and inject falsified data into the sensors. 

In fact, there have been numerous reported attacks and vulnerabilities of smart inverters in the news and literature. For example, the Dutch Government Inspectorate for Digital Infrastructure (RDI) has recently reported that none of the nine examined solar inverters met the requirements for cybersecurity, as they can be hacked, remotely disabled or exploited for DDoS attacks \cite{van2023eenmalige}. The European Commission released a report on the cybersecurity and resilience of Europe’s telecommunications and electricity sectors, which highlights the concerns of supply chain security risks to enable concerted attacks on the EU's energy infrastructure \cite{EU_cyberrisk}.

The US Cyber and Infrastructure Security Agency (CISA) issued two warnings relevant to inverter products from manufacturer Enphase: Envoy communication gateway and Installer Toolkit Android app, which contain security bugs and hardcoded credentials that could allow an attacker to gain root access to the affected products \cite{CISA_enphase}. NIST examined several known smart inverter vulnerabilities documented in the National Vulnerability Database (NVD) and tested five smart inverters for vulnerabilities. They provided guidelines to smart inverter manufacturers to improve the cybersecurity capabilities needed in their products \cite{NIST_smartinverter}.

The Dutch Institute for Vulnerability Disclosure (DIVD) researchers also uncovered six zero-day vulnerabilities in Enphase IQ Gateway devices \cite{DIVD_enphase}. Bitdefender found that the smart inverter Solarman platform’s API contained a flaw that could allow attackers to generate authorization tokens for any account on the platform, allowing them to control the smart inverters remotely \cite{Bitdefender}. Note that Advanced Persistent Threat actors are increasingly targeting at decentralized energy systems and smart inverters \cite{BCELH25APT}.

\aptLtoX[graphic=no,type=html]{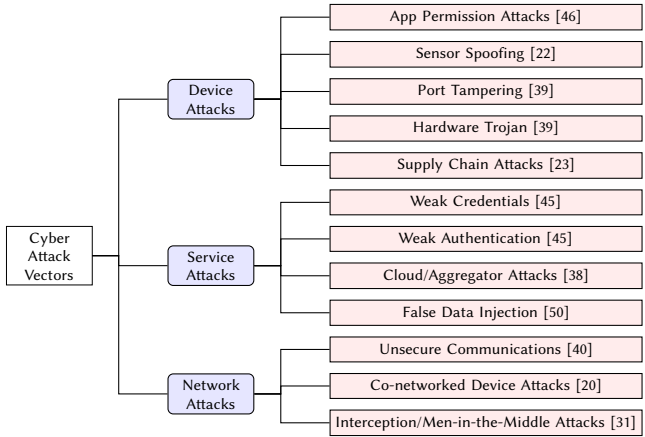
\begin{figure}[t!]
\begin{minipage}{0.50\textwidth}

\centering
    
% using only the basic, xnode, and tnode styles
\tikzset{
    basic/.style  = {draw, text width=1cm, align=center, font=\sffamily, rectangle, line width=0},
    root/.style   = {basic, text width=1cm, rounded corners=2pt, thin, align=center, fill=green!30},
    onode/.style = {basic, thin, rounded corners=2pt, align=center, fill=green!60,text width=1cm,},
    tnode/.style = {basic, thin, align=left, fill=pink!30, text width=19em, align=center},
    xnode/.style = {basic, thin, rounded corners=2pt, align=center, fill=blue!10,text width=1cm,},
    wnode/.style = {basic, thin, align=left, fill=pink!10!blue!80!red!10, text width=1cm},
    edge from parent/.style={draw=black, edge from parent fork right}

}

{\scriptsize

\begin{forest} for tree={
    grow=east,
    growth parent anchor=west,
    parent anchor=east,
    child anchor=west,
    edge path={\noexpand\path[\forestoption{edge}, >={latex}] 
         (!u.parent anchor) -- +(10pt,0pt) |-  (.child anchor) 
         \forestoption{edge label};}
}
% l sep is used for arrow distance
[Cyber Attack Vectors, basic,  l sep=10mm,
    [Network Attacks, xnode,  l sep=10mm,
        [Interception/Men-in-the-Middle Attacks \cite{Jan2019Toward}, tnode]		
        [Co-networked Device Attacks \cite{Coppolino2015My}, tnode] 
		[Unsecure Communications \cite{Mechev2020Cybersecurity}, tnode] 
	]
    [Service Attacks, xnode,  l sep=10mm,
        [False Data Injection \cite{Wang2017A}, tnode] 
		[Cloud/Aggregator Attacks \cite{Liu2017Cyber}, tnode] 
        [Weak Authentication \cite{Shah2019A}, tnode]
		[Weak Credentials \cite{Shah2019A}, tnode]
    ]	
    [Device Attacks, xnode,  l sep=10mm,
        [Supply Chain Attacks \cite{Duman2019Modeling}, tnode]	
        [Hardware Trojan \cite{Manzanares2005Attacks}, tnode]
        [Port Tampering \cite{Manzanares2005Attacks}, tnode]
        [Sensor Spoofing \cite{Delarea2022Practical}, tnode]  
		[App Permission Attacks \cite{Sivaraman2016Smart-Phones}, tnode]
	]	
]
\end{forest}
}
\vspace{-5pt}\caption{Taxonomy of cyber attack vectors on smart inverters.} \label{fig:cyberatkvec}
\end{minipage}
\end{figure}
\begin{figure}
\begin{minipage}{0.47\textwidth}

\centering
    
% using only the basic, xnode, and tnode styles
\tikzset{
    basic/.style  = {draw, text width=1cm, fill=gray!30, align=center, font=\sffamily, rectangle, line width=0},
    root/.style   = {basic, text width=1cm, rounded corners=2.2pt, thin, align=center, fill=green!30},
    onode/.style = {basic, thin, rounded corners=2pt, align=center, fill=green!60,text width=1cm,},
    tnode/.style = {basic, thin, align=left, fill=yellow!10, text width=15em, align=center},
    xnode/.style = {basic, thin, rounded corners=2pt, align=center, fill=green!20,text width=1.5cm,},
    wnode/.style = {basic, thin, align=left, fill=pink!10!blue!80!red!10, text width=1cm},
    edge from parent/.style={draw=black, edge from parent fork right}

}

{\scriptsize

\begin{forest} for tree={
    grow=east,
    growth parent anchor=west,
    parent anchor=east,
    child anchor=west,
    edge path={\noexpand\path[\forestoption{edge}, >={latex}] 
         (!u.parent anchor) -- +(10pt,0pt) |-  (.child anchor) 
         \forestoption{edge label};}
}
% l sep is used for arrow distance
[Grid Attack Vectors, basic,  l sep=10mm,
    [Reactive Power Control Attacks, xnode,  l sep=10mm,
        [Comm. Interface Exploits \cite{Jones2023Artificial}, tnode]
        [Data Poisoning \cite{Jones2023Artificial}, tnode]
        [Control Manipulation \cite{Kandasamy2020Prosumer}, tnode]		
    ]	
    [Voltage Control Attacks, xnode,  l sep=10mm,
        [Comm. Interface Exploits \cite{Fard2020Holistic}, tnode]
        [Data Poisoning \cite{Ahmadzadeh2022Detection}, tnode]
        [Control Manipulation \cite{Fard2020Holistic}, tnode]		
	]
    [Frequency Control Attacks, xnode,  l sep=10mm,
        [Comm. Interface Exploits \cite{Wu2018Resonance}, tnode]
        [Data Poisoning \cite{Abbaspour2020Resilient}, tnode]
        [Control Manipulation \cite{Wu2018Resonance}, tnode]	 		
    ]	
]
\end{forest}
}
\vspace{-5pt}\caption{Taxonomy of grid attack vectors by smart inverters.} \label{fig:gridatkvec}
\end{minipage}
\end{figure}
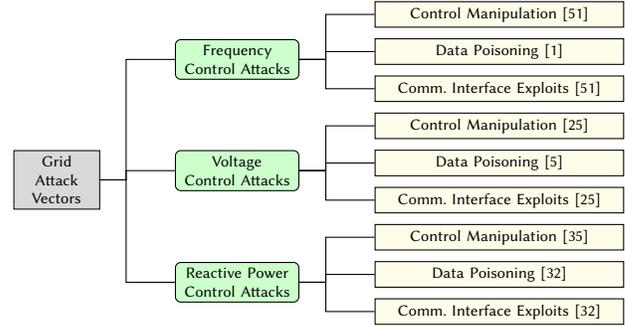
}{
\begin{figure*}[t!]
\begin{minipage}{0.52\textwidth}

\centering
    
% using only the basic, xnode, and tnode styles
\tikzset{
    basic/.style  = {draw, text width=1cm, align=center, font=\sffamily, rectangle, line width=0},
    root/.style   = {basic, text width=1cm, rounded corners=2pt, thin, align=center, fill=green!30},
    onode/.style = {basic, thin, rounded corners=2pt, align=center, fill=green!60,text width=1cm,},
    tnode/.style = {basic, thin, align=left, fill=pink!30, text width=19em, align=center},
    xnode/.style = {basic, thin, rounded corners=2pt, align=center, fill=blue!10,text width=1cm,},
    wnode/.style = {basic, thin, align=left, fill=pink!10!blue!80!red!10, text width=1cm},
    edge from parent/.style={draw=black, edge from parent fork right}

}

{\scriptsize

\begin{forest} for tree={
    grow=east,
    growth parent anchor=west,
    parent anchor=east,
    child anchor=west,
    edge path={\noexpand\path[\forestoption{edge}, >={latex}] 
         (!u.parent anchor) -- +(10pt,0pt) |-  (.child anchor) 
         \forestoption{edge label};}
}
% l sep is used for arrow distance
[Cyber Attack Vectors, basic,  l sep=10mm,
    [Network Attacks, xnode,  l sep=10mm,
        [Interception/Men-in-the-Middle Attacks \cite{Jan2019Toward}, tnode]		
        [Co-networked Device Attacks \cite{Coppolino2015My}, tnode] 
		[Unsecure Communications \cite{Mechev2020Cybersecurity}, tnode] 
	]
    [Service Attacks, xnode,  l sep=10mm,
        [False Data Injection \cite{Wang2017A}, tnode] 
		[Cloud/Aggregator Attacks \cite{Liu2017Cyber}, tnode] 
        [Weak Authentication \cite{Shah2019A}, tnode]
		[Weak Credentials \cite{Shah2019A}, tnode]
    ]	
    [Device Attacks, xnode,  l sep=10mm,
        [Supply Chain Attacks \cite{Duman2019Modeling}, tnode]	
        [Hardware Trojan \cite{Manzanares2005Attacks}, tnode]
        [Port Tampering \cite{Manzanares2005Attacks}, tnode]
        [Sensor Spoofing \cite{Delarea2022Practical}, tnode]  
		[App Permission Attacks \cite{Sivaraman2016Smart-Phones}, tnode]
	]	
]
\end{forest}
}
\vspace{-5pt}\caption{Taxonomy of cyber attack vectors on smart inverters.} \label{fig:cyberatkvec}
\end{minipage}
\begin{minipage}{0.47\textwidth}

\centering
    
% using only the basic, xnode, and tnode styles
\tikzset{
    basic/.style  = {draw, text width=1cm, fill=gray!30, align=center, font=\sffamily, rectangle, line width=0},
    root/.style   = {basic, text width=1cm, rounded corners=2.2pt, thin, align=center, fill=green!30},
    onode/.style = {basic, thin, rounded corners=2pt, align=center, fill=green!60,text width=1cm,},
    tnode/.style = {basic, thin, align=left, fill=yellow!10, text width=15em, align=center},
    xnode/.style = {basic, thin, rounded corners=2pt, align=center, fill=green!20,text width=1.5cm,},
    wnode/.style = {basic, thin, align=left, fill=pink!10!blue!80!red!10, text width=1cm},
    edge from parent/.style={draw=black, edge from parent fork right}

}

{\scriptsize

\begin{forest} for tree={
    grow=east,
    growth parent anchor=west,
    parent anchor=east,
    child anchor=west,
    edge path={\noexpand\path[\forestoption{edge}, >={latex}] 
         (!u.parent anchor) -- +(10pt,0pt) |-  (.child anchor) 
         \forestoption{edge label};}
}
% l sep is used for arrow distance
[Grid Attack Vectors, basic,  l sep=10mm,
    [Reactive Power Control Attacks, xnode,  l sep=10mm,
        [Comm. Interface Exploits \cite{Jones2023Artificial}, tnode]
        [Data Poisoning \cite{Jones2023Artificial}, tnode]
        [Control Manipulation \cite{Kandasamy2020Prosumer}, tnode]		
    ]	
    [Voltage Control Attacks, xnode,  l sep=10mm,
        [Comm. Interface Exploits \cite{Fard2020Holistic}, tnode]
        [Data Poisoning \cite{Ahmadzadeh2022Detection}, tnode]
        [Control Manipulation \cite{Fard2020Holistic}, tnode]		
	]
    [Frequency Control Attacks, xnode,  l sep=10mm,
        [Comm. Interface Exploits \cite{Wu2018Resonance}, tnode]
        [Data Poisoning \cite{Abbaspour2020Resilient}, tnode]
        [Control Manipulation \cite{Wu2018Resonance}, tnode]	 		
    ]	
]
\end{forest}
}
\vspace{-5pt}\caption{Taxonomy of grid attack vectors by smart inverters.} \label{fig:gridatkvec}
\end{minipage}
\end{figure*}}

\vspace{-5pt}
\subsection{Attack Vectors of Smart Inverters} \label{sec:related}

This section reviews the known attack vectors on smart inverters and how to leverage smart inverters to attack the power grid.

\smallskip
\noindent
$\triangleright$ {\bf Cyber Attack Vectors on Smart Inverters:} We classify the cyber attack vectors on smart inverters, including those documented in Common Vulnerabilities and Exposures (CVEs), as follows:
\begin{enumerate}[leftmargin=*]

\item {\em Device Attacks:} These attacks exploit the security flaws in the physical hardware and embedded software of smart inverters. Key examples include app permission attacks, sensor spoofing, port tampering, hardware trojans, supply chain attacks by embedding malicious code or backdoors through manufacturers.

\item {\em Service Attacks:} These attacks exploit the weaknesses in the services that manage and interact with smart inverters. Key examples include weak credentials (e.g., hard-coded/weak passwords), weak authentication, cloud/aggregator attacks that target the remote control parties and false data injection.

\item {\em Network Attacks:} These attacks exploit the vulnerabilities in the networks and communications with smart inverters. Key examples include unsecure communications, network infiltration through co-networked smart devices and DOS attacks. 

\end{enumerate}

%\smallskip
\noindent
$\triangleright$ {\bf Grid Attack Vectors by Smart Inverters:} Once smart inverters are compromised, they can be used for grid-level attacks to disrupt the power grid. We classify the grid attack vectors as follows:
\begin{enumerate}[leftmargin=*]

\item {\em Frequency Control Attacks:} Keeping the grid frequency stable is a crucial balancing act. In Australia, the grid frequency needs to stay close to 50Hz. Any deviation away from this can lead to cascading power system failures, such as wide-wide power outages. Frequency Co-optimised Essential System Services (FCESS) \cite{lal2021essential} or Frequency Control Ancillary Services (FCAS) are essential for controlling the grid frequency by balancing the generation and load in real-time. Frequency control attacks may use compromised smart inverters to disrupt frequency control and to cause wide-scale power instability.

\item {\em Voltage Control Attacks:} It is important to ensure that the voltage levels remain within safe operating limits, preventing issues such as overvoltage and undervoltage, which can damage equipment and disrupt the power supply. Voltage control attacks exploit specific vulnerabilities of consumer energy systems, targeting the mechanisms responsible for maintaining voltage levels. Smart inverters support low-voltage ride-through, which can be exploited for voltage control attacks. 

\item {\em Reactive Power Control Attacks:} Reactive power is generated from capacitive elements or inductive elements of power systems. Proper reactive levels are necessary to certain equipment. Reactive power control attacks can disrupt the reactive power levels, leading to voltage instability and equipment damage. Smart inverters also support reactive power compensation, which can be exploited for reactive power control attacks. 

\end{enumerate}
Note that frequency control attacks can lead to wide-scale instability, whereas voltage control and reactive power control attacks often lead to local instability. Hence, this paper focuses on frequency control attacks by smart inverters. These attacks can be triggered by manipulation of control units, data poisoning to control units, or exploiting communication interfaces. 

Figs.~\ref{fig:cyberatkvec}-\ref{fig:gridatkvec} present the taxonomies of the cyber attack vectors on smart inverters and the grid attack vectors by smart inverters, along with citations of the related work in the literature.

\vspace{-5pt}
\subsection{Power System Requirements and Measures} 

Modern power grids are designed to operate to satisfy the following power system requirements in principle:

\begin{itemize}[leftmargin=*]

\item {\em Power System Reliability}: A power system should offer sufficient generation, demand response and network capacity to satisfy consumers' demands with substantial confidence.

\item {\em Power System Security}: A power system should ensure that the operational system parameters, such as frequency and voltage, are properly maintained within the defined limits.

\end{itemize}
Power grid operators are expected to address the balance of generation and load in real-time, thus meeting power system reliability and security. However, unexpected scenarios occasionally happen in practice that may jeopardize power system security.
There are two measures that grid operators can cope with unexpected scenarios:

\smallskip
\noindent
$\triangleright$ {\bf Contingency Mechanisms:} 
Grid operators specify a set of {\em credible contingency events} \cite{AEMO_creditcont} and ensure proper mechanisms (e.g., frequency control) in response to these events. Typical credible contingency events include natural catastrophic events (e.g., bushfires, storms, floods), power system failures (e.g., short-circuit, transmission wire tripping) and severe events like the loss of a major power plant. Also, the creditable contingency events will vary dynamically in response to generation dispatch within a risk level. 

However, the current lists of creditable contingency events for most grid operators often do not consider malicious actors to inflict adverse impacts on the grid (e.g., cyberattacks). Note that savvy attackers may be able to exploit certain weaknesses in the contingency capacity mechanisms and orchestrate their attacks to take advantage of the most vulnerable opportunities. Attacks on the power grid may cause long-term outages to affect power system reliability, or disruption to the operational system parameters (frequency, voltage) by short-term fluctuations of generation or load. In this paper, we provide a realistic assessment of the plausibility and impacts of wide-scale power instability caused by savvy attackers, with explicit consideration of contingency capacity mechanisms in operation for power system security.

\smallskip
\noindent
$\triangleright$ {\bf System Inertia:} Another way to enhance systems security is to slow down the impacts caused by unexpected scenarios. Power grids, like many dynamical systems, possess system inertia to resist the change of the operating states by the extractable kinetic energy, as manifested by the rotating masses in the grid (e.g., synchronous generation, flywheels, standby generators). Grid operators maintain a certain level of system inertia in the grid to allow sufficient time to activate further generation in the case of unexpected scenarios.

\vspace{-5pt}
\subsection{Assessment Studies of Cyberattacks on Grid}

Although frequency control attacks have been studied via small-scale simulations (e.g., \cite{Wu2018Resonance,CSCD24}),  there are recent studies based on real-world grid dataset to assess the plausibility and impacts of power instability caused by cyberattacks on DERs (e.g., smart inverters, EV charging stations) . In Table~\ref{tab:related_work}, we present a comparison of these assessment studies with this work. Acharya et al. \cite{ADK20ev} evaluated cyberattacks via EV charging stations in NYC.  Goerke et al. \cite{goerke2024controls} estimated the portion of compromised DERs to overwhelm the frequency containment reserve capacity (4500 MW) of ENTSO-E (European Network of Trans. System Operators for Electricity). Musleh et al. \cite{musleh2024experimental} estimated the costs of reserve capacity to compensate for the loss of inverter-based generation in South Australia. 

But there are shortcomings in these prior studies. First, \cite{ADK20ev,goerke2024controls,musleh2024experimental} do not consider the dynamic nature of contingency-based frequency control. They consider either static contingency capacity or no contingency service. Second, system inertia is integral to a complete understanding of frequency deviation, as it shows the timescale of frequency response to attacks. None of these prior studies evaluate system inertia in practice. Hence, these studies may overestimate the impacts of attacks, if there is sufficient system inertia in the grid to counteract the effects of imbalance of generation and load. 

The novelty of this study is to provide a realistic assessment of the plausibility and impacts of wide-scale power instability caused by cyberattacks on smart inverters, with explicit consideration of real-world contingency services and system inertia modelling and real-life electricity market data. Our insights are more accurate and holistic than the prior studies, and offer insightful findings to improve power security with high penetration of DPV.

\begin{table}[t!]
\centering
\scriptsize\tabcolsep4pt
\begin{tabularx}{1.05\columnwidth}{@{}p{0.35cm}|X|X@{}}
\hline
 & \textbf{Attack Scenarios} & \textbf{Results \& Remarks} \\
 \hline
  \vspace{10pt}
    \textbf{\cite{ADK20ev}} & 
\vspace{-6pt}

\begin{itemize}\leftskip-15pt%[leftmargin=*] 
    \item Aggregate high-wattage load from EV charging stations to cause frequency instability in the grid
    \item Based on data-driven attack strategy with knowledge of eigenvalues of the grid model 
\end{itemize} & 
\vspace{-6pt}

\begin{itemize}\leftskip-15pt%[leftmargin=*] 
    \item Evaluated on public data of EV charging stations in NYC and power grid data from US Energy and NY ISO 
    \item Simulations show this attack is
infeasible at the current penetration of EVs 
    \item {\bf\em No contingency-based frequency control or system inertia is considered} \vspace{-6pt}
\end{itemize} \\ \hline 
  \vspace{10pt}
    \textbf{\cite{goerke2024controls}} &  
\vspace{-6pt}

\begin{itemize}\leftskip-15pt%[leftmargin=*] 
    \item Aggregate DERs (EVs, inverters, building heat pumps, battery energy storage) to cause frequency instability in the grid
    \item Overwhelm the capacity of frequency containment reserve
\end{itemize} & 
\vspace{-6pt}

\begin{itemize}\leftskip-15pt%[leftmargin=*] 
    \item Estimate the portion of DERs to overwhelm the frequency containment reserve capacity of ENTSO-E (European Network of Trans. System Operators for Electricity)
    \item {\bf\em Only static contingency capacity is considered, no system inertia is modeled}  \vspace{-10pt}  
\end{itemize} \\ \hline
  \vspace{8pt}
    \textbf{\cite{musleh2024experimental}} &  
\vspace{-6pt}

\begin{itemize}\leftskip-15pt%[leftmargin=*] 
    \item Disrupt voltage and frequency ride-through capabilities of smart inverters
    \item Manipulate control in smart inverters to destabilize energy supplies to the grid
\end{itemize} & 
\vspace{-6pt}

\begin{itemize}\leftskip-15pt%[leftmargin=*] 
    \item Estimate the costs of reserve capacity to compensate the loss of inverter-based generation in South Australia
    \item {\bf\em No contingency-based frequency control or system inertia is considered} \vspace{-10pt} 
\end{itemize} \\ \hline
  \vspace{8pt}
\textbf{This work} &  
\vspace{-6pt}

\begin{itemize}\leftskip-15pt%[leftmargin=*] 
    \item Disrupt smart inverters to cause sudden generation loss or hike to create frequency instability in the grid  
    \item Optimize attacks to the contingency capacity mechanism based on open electricity market data
\end{itemize} & 
\vspace{-6pt}

\begin{itemize}\leftskip-15pt%[leftmargin=*] 
    \item  Assess realistic attack scenarios and opportunities, and the impacts based on electricity market data in Australia
    \item Devise a predictive model for optimized attacks on frequency control
    \item {\bf\em Consider real-world contingency-based frequency control and system inertia} \vspace{-15pt}  
\end{itemize} \\
\hline 
\end{tabularx}
\caption{Comparison of assessment studies of power grid by cyberattacks on consumer energy resources.}
\label{tab:related_work}  \vspace{-25pt}
\end{table}

\aptLtoX[graphic=no,type=html]{\begin{figure}[th!]
\begin{minipage}{0.45\textwidth}
\centering
\includegraphics[width=0.9\columnwidth]{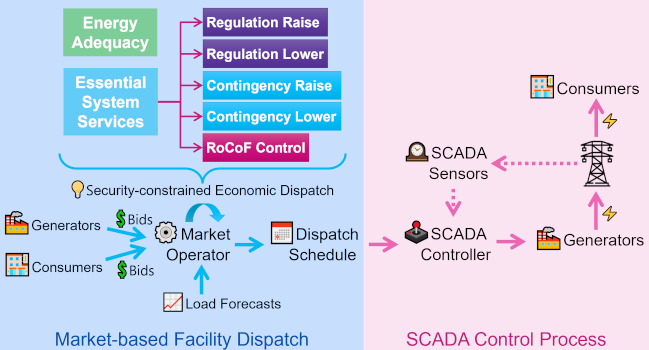}  \vspace{-10pt}
\caption{Procedures of Australia's electricity markets.} 
\label{fig:marketflow} \vspace{-5pt}
\end{minipage}
\end{figure}
\begin{figure}
\begin{minipage}{0.45\textwidth}
\centering
\includegraphics[width=1\columnwidth]{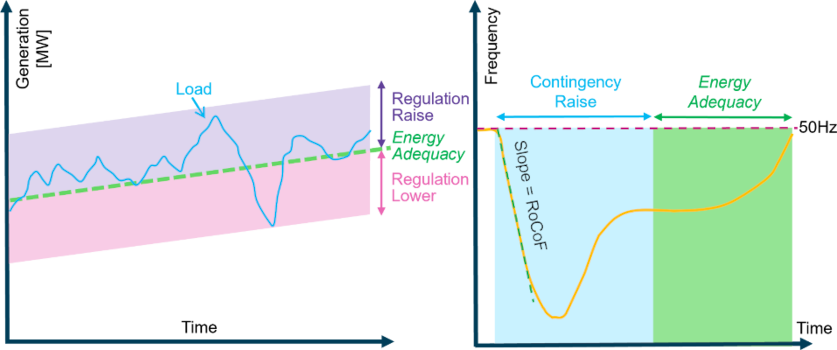}  %\vspace{-5pt}
\caption{Frequency control by ESS (Regulation Raise/Lower, Contingency Raise/Lower, RoCoF control).} 
\label{fig:marketcontrol} \vspace{-5pt}
\end{minipage}
\end{figure}}{\begin{figure*}[th!]
\begin{minipage}{0.45\textwidth}
\centering
\includegraphics[width=0.9\columnwidth]{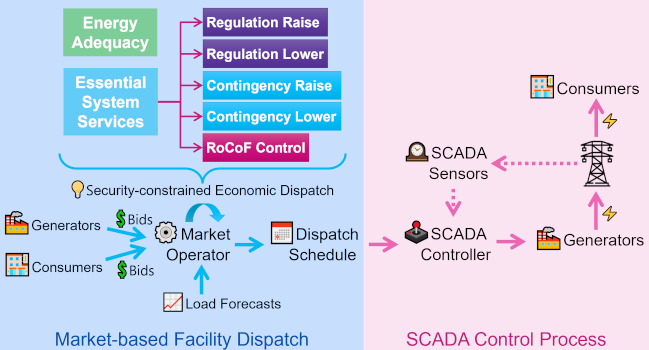}  \vspace{-10pt}
\caption{Procedures of Australia's electricity markets.} 
\label{fig:marketflow} \vspace{-5pt}
\end{minipage}
\begin{minipage}{0.45\textwidth}
\centering
\includegraphics[width=1\columnwidth]{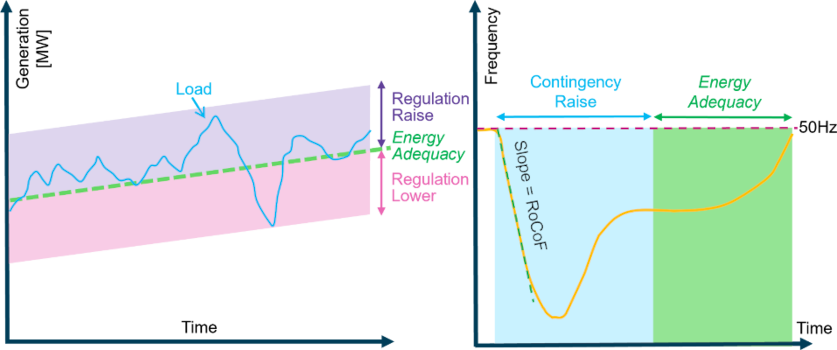}  %\vspace{-5pt}
\caption{Frequency control by ESS (Regulation Raise/Lower, Contingency Raise/Lower, RoCoF control).} 
\label{fig:marketcontrol} \vspace{-5pt}
\end{minipage}
\end{figure*}}

\vspace{-5pt}
\section{Australia's Electricity Markets} \label{sec:wem}

In this section, we provide an overview of Australia's electricity market for our case study. We note that other places (e.g., California, Texas, UK and Ireland) have also adopted a similar electricity market structure in some aspects. The Australia's electricity grid systems are divided into two networks – one interconnecting six eastern states facilitated by the National Electricity Market (NEM) and another interconnecting only Western Australia facilitated by the Wholesale Electricity Market (WEM). Both markets are operated by the Australian Energy Market Operator (AEMO). This paper focuses on one of the states in Australia. 

The primary goal of an electricity market is to satisfy power system reliability by balancing generation and load at open-market prices. The secondary goal is to ensure power system security by securing sufficient standby capacity in case of fluctuations of load, shortage of renewable energy, and emergencies of generator outages. In Australia's power grid, both power system reliability and security are attained through a single co-optimized dispatch process. Australia's electricity markets provide not only open market datasets but also transparent specifications of the contingency capacity mechanism, which allows us to assess various cyberattack scenarios.

\vspace{-5pt}
\subsection{Procedures of Australia's Electricity Markets} 

We briefly overview the procedures of Australia's electricity markets \cite{WEM_rules}, which are summarized by the following stages (and illustrated in Fig.~\ref{fig:marketflow}):
\begin{enumerate}[leftmargin=*]

\item {\bf Market-based Co-optimised Facility Dispatch}. In every 5-minute interval, the market participants (i.e., generators, retail consumers) submit bids of prices and quantities based on cost estimation and facility availability. The market operator optimizes a security-constrained economic dispatch to produce a {\em dispatch schedule} that matches load forecasts at the lowest cost, subject to power system requirements. The co-optimised dispatch schedule consists of two parts: (1) {\em energy adequacy} of generator facilities to match load forecasts, (2) {\em essential system services} for frequency control to maintain stable grid frequency.

\begin{enumerate}[leftmargin=*]

\item {\bf Load Forecasts}: The market operator forecasts the demand in the short-term (one-week) and medium-term (multi-month). This helps to guide the dispatch process. There will be likely fluctuations of load or generation from renewable energy in operation (particularly, due to distributed PV). The market operator rectifies the balance between generation and load by essential system services.

\item {\bf Security-constrained Economic Dispatch}: The objective of a security-constrained economic dispatch is to dispatch facilities for both energy adequacy and essential system services together with the lowest total cost for generation and frequency control. The market operator adjusts the power system constraints according to energy adequacy in the economic dispatch to reflect the power system security requirement levels (e.g., taking a possible failure of a generator into account).

\end{enumerate}

\item {\bf SCADA Control Process}: Based on a dispatch schedule, the market operator monitors and controls the generation facilities in operation through the Supervisory Control and Data Acquisition (SCADA) system. The market operator gathers real-time sensor data of the operating states of the grid via SCADA sensors and dynamically regulates individual generation facilities to follow the dispatch schedule. In case of deviation of operating states from the desirable range (e.g., the grid frequency deviates from 50Hz), the market operator intervenes through a sequence of remedies - first by invoking the essential system services and, if not effective, then by controlled load/generation shedding.

\end{enumerate}

\begin{figure*}[t!]
\vspace{-5pt}
\includegraphics[width=1\textwidth]{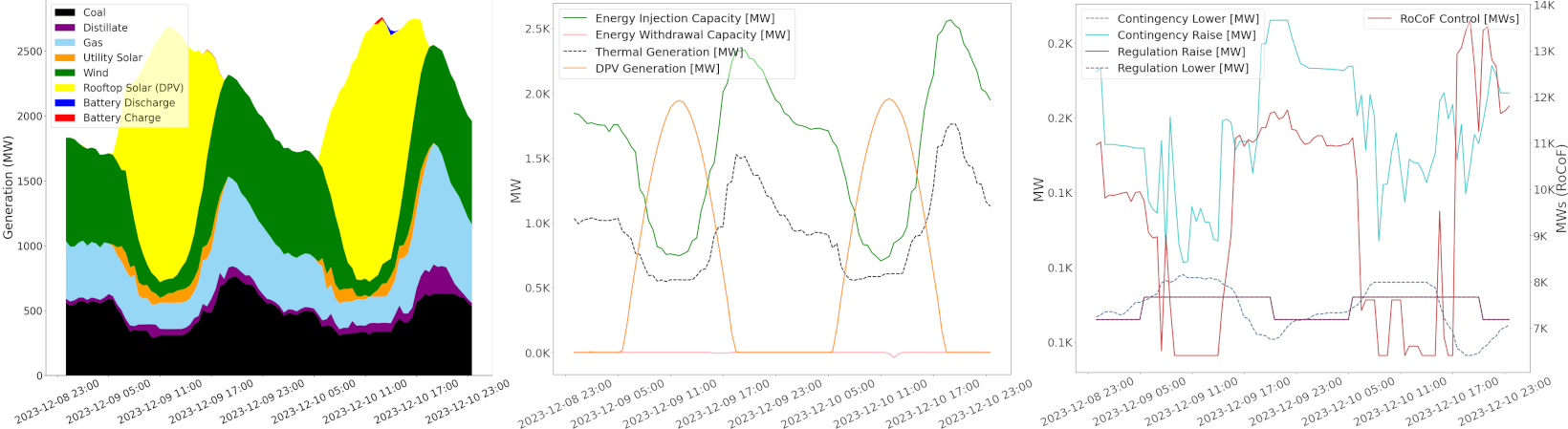}  \vspace{-20pt}
\caption{Two-day market data from one of the Australian states on 8 Dec 23 - 10 Dec 23: (Left) The generation mix. (Middle) The dispatch schedule of energy and essential system services. (Right) The market prices of energy and essential system services.} 
\label{fig:wem-day-data} \vspace{-5pt}
\end{figure*}

\vspace{-5pt}
\subsection{Essential System Services (ESS)} 

Frequency Co-optimised ``Essential System Services'' (ESS) \cite{lal2021essential}, also known as ancillary services, ensures power system security. The core of ESS is to enable frequency control. If there is an imbalance between generation and load, the grid frequency will deviate from the standard, leading to instability or, in the worst case, cascading failure and blackouts\footnote{The wide-scale power outage in Spain and Portugal on 28 April 2025 was reportedly caused by abnormal grid frequency deviation \cite{SpainPortOutage}.}. To maintain the grid frequency within the desirable range around 50Hz, the market operator relies on ESS to provide a rapid injection to or a reduction of energy from the grid. 

Unexpected variations in generation occur from time to time. For example, on 16 March 2021, a fast-moving cloudbank in wide areas in Western Australia resulted in decreased rooftop solar generation by around 300 MW within a few minutes and the grid frequency dropped to 49.5Hz, which is generally associated with the trip of a large generator. ESS is necessary to ensure rapid corrections of generation and load to mitigate the grid frequency deviation. 

In general, frequency co-optimised ESS in Australia's power grid are divided into the following classes:
\begin{itemize}[leftmargin=*]

\item {\bf Regulation} aims to regulate the generation or load of a facility in response to {\em minor} fluctuations of generation and load, by frequently adjusting its energy injection of generation or withdrawal of load in accordance with the market operator's control scheme. The regulation facilities that are dispatched in the dispatch schedule will be readily triggered by the market operator via SCADA control. Regulation service consists of raise and lower sub-services:
(1) {\bf Regulation Raise} aims to raise the frequency by increasing generation or decreasing load from the energy adequacy level. 
(2) {\bf Regulation Lower} aims to lower the frequency by decreasing generation or increasing load from the energy adequacy level. 
Note that regulation facilities need to be accredited for the capacity that can be provided in regulation raise and lower services. 

\item {\bf Contingency} aims to provide the standby capability of a facility in reserve so that it can rapidly adjust its energy injection or withdrawal in response to a credible contingency event (e.g., a failure of a generator in the dispatch schedule). Contingency service can respond to {\em larger} fluctuations of generation and load than regulation services occasionally. Contingency service consists of raise and lower sub-services: (1) {\bf Contingency Raise} aims to raise the frequency by increasing generation or decreasing load for a credible contingency event. 
(2) {\bf Contingency Lower} aims to lower the frequency by decreasing generation or increasing load for a credible contingency event.

\item {\bf RoCoF Control} aims to provide system inertia in the power grid to mitigate instantaneous changes of frequency. System inertia is the kinetic energy that is extractable from the rotating mass of a synchronous machine coupled to the power grid (e.g., a gas turbine generator). The amount of inertia presented in the power grid is inversely proportional to the rate of change of frequency (RoCoF) \cite{AM90inertia}. Ensuring sufficient system inertia can slow down frequency deviation in case of a mismatch between generation and load, offering more time to activate the contingency facilities. 

\end{itemize}

We illustrate the basic idea of frequency control by ESS in Fig.~\ref{fig:marketcontrol}. Regulation service is supposed to rectify minor deviations from the energy adequacy level. In case of a larger deviation, contingency service is activated to bring the grid frequency to a stable level. RoCoF control ensures a mild rate of change of frequency in the grid before contingency service can be activated.

We plot two-day data of the dispatch schedule from one of the Australian states on 8 Dec 23 - 10 Dec 23 in Fig~\ref{fig:wem-day-data}. It shows that distributed PV (DPV) provides the highest generation percentage at noon time, while the capacity for regulation raise and RoCoF control is at the lowest. It also shows the market price fluctuations of ESS during the period.

\smallskip
\noindent
$\triangleright$ {\bf Emergency Mechanisms:} 
In case that ESS is insufficient in correcting the frequency deviation, further emergency mechanisms will be triggered to restore the desirable grid frequency to prevent cascading power failures and wide-scale outages as follows:

\begin{itemize}[leftmargin=*]

\item {\bf\em UFLS} (Under Frequency Load Shedding): This is an automatic scheme \cite{aemo_ufls} when the grid frequency is substantially lower than 50Hz in a major unexpected event that cannot be mitigated by contingency service. This scheme is divided into 5 stages as described in Table~\ref{tab:ufls}, and each stage has a set of prioritization of load to shed and a reaction time of 0.4sec.

\begin{table}[h!]
\centering %\vspace{-5pt}
\resizebox{0.7\columnwidth}{!}{%
%{\scriptsize
\begin{tabular}{@{}ccc@{}}
\hline
Stage & Initiation Threshold (Hz) & Load Shed Quantity (\%) \\
\hline
1 & 48.75 & 15 \\
2 & 48.50 & 15 \\
3 & 48.25 & 15 \\
4 & 48.00 & 15 \\
5 & 47.75 & 15 \\
\hline
\end{tabular}
}
\caption{Under frequency load shedding scheme (UFLS).} \label{tab:ufls} \vspace{-20pt}
\end{table}

\item {\bf\em OFGS} (Over Frequency Generation Shedding): When the grid frequency is substantially higher than 50Hz, some generators will be forced to have emergency disconnection, which may lead to adverse physical effects on the generators.

\end{itemize}

\aptLtoX[graphic=no,type=html]{\begin{figure}[t!]
\begin{minipage}{0.4\textwidth}
\centering
\includegraphics[scale=1.07]{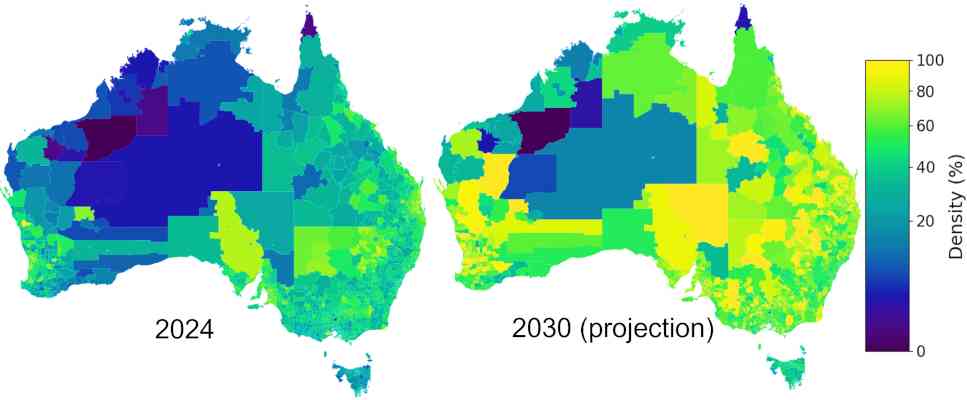}  \vspace{-10pt}
\caption{Current and projected residential DPV density in Australia.} 
\label{fig:wa-pv-density} %\vspace{-15pt}
\end{minipage}
\end{figure}
\begin{figure}
\begin{minipage}{0.3\textwidth}
\centering \vspace{-20pt}
\includegraphics[scale=0.1]{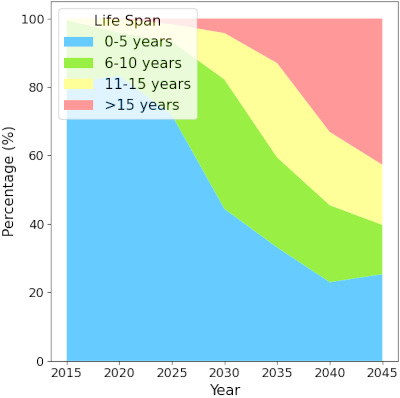}  \vspace{-10pt}
\caption{Projected lifespan distribution of smart inverters in Australia.} 
\label{fig:lifespan} %\vspace{-15pt}
\end{minipage} \quad
\end{figure}
\begin{figure}
\begin{minipage}{0.22\textwidth}
\centering
\includegraphics[scale=1.22]{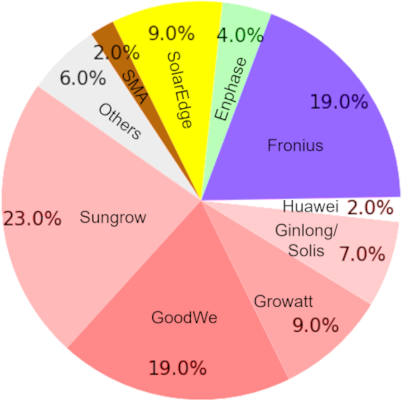}  \vspace{-5pt}
\caption{Market share of smart inverters in Australia.} 
\label{fig:marketshare} %\vspace{-15pt}
\end{minipage}
\end{figure}
}{
\begin{figure*}[t!]
\begin{minipage}{0.4\textwidth}
\centering
\includegraphics[scale=0.22]{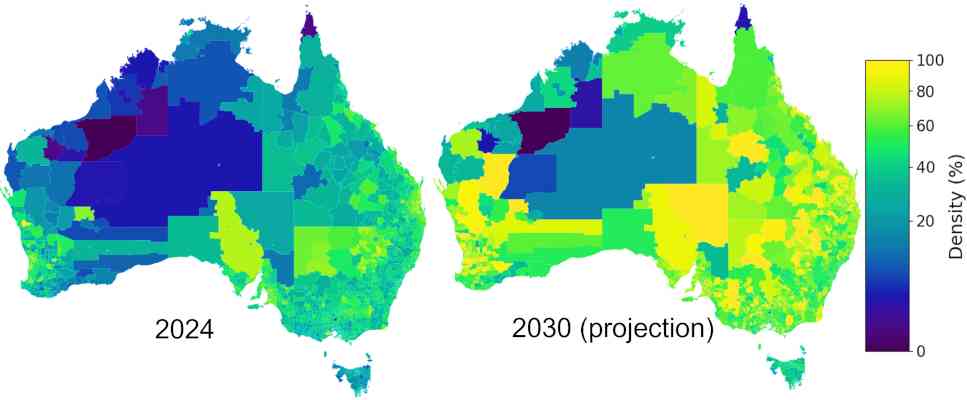}  \vspace{-10pt}
\caption{Current and projected residential DPV density in Australia.} 
\label{fig:wa-pv-density} %\vspace{-15pt}
\end{minipage} \quad
\begin{minipage}{0.3\textwidth}
\centering
\includegraphics[scale=0.25]{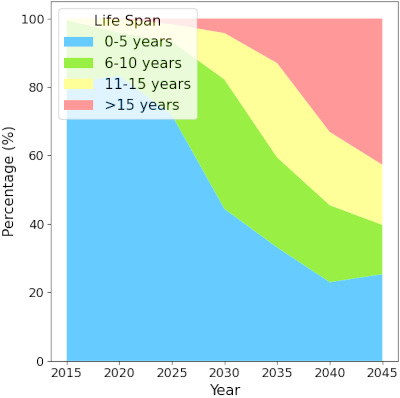}  \vspace{-10pt}
\caption{Projected lifespan distribution of smart inverters in Australia.} 
\label{fig:lifespan} %\vspace{-15pt}
\end{minipage} \quad
\begin{minipage}{0.22\textwidth}
\centering
\includegraphics[scale=0.23]{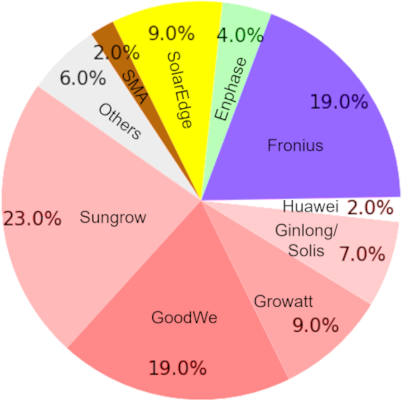}  \vspace{-5pt}
\caption{Market share of smart inverters in Australia.} 
\label{fig:marketshare} %\vspace{-15pt}
\end{minipage}
\end{figure*}}
\vspace{-5pt}

\subsection{ESS Facilities} \label{sec:facility}

We take a closer look at the active facilities that offer ESS in an unnamed Australian state. In Table~\ref{tab:facilities} in Appendix~\ref{apx:facilities}, we list the data of active ESS facilities in that state in \cite{aemo_fcess_accreditation}. The table includes details on the generator types and the associated maximum capacities in MW for different ESS classes. The table highlights the utilization rates of these facilities from Oct 2023 to Sept 2024. 

Two properties reflect the significance of each facility to respond to contingency events:
\begin{itemize}[leftmargin=*]

\item {\em Speed Factor} represents how quickly a facility responds to frequency changes, with the acceptable range typically between 0.2 and 15 secs. A lower speed factor indicates a faster response to frequency deviations, meaning these facilities are considered primary responders during emergencies. 

\item {\em RoCoF Ride-Through} refers to a facility's ability to withstand sudden frequency changes without disconnecting from the grid. Facilities with higher RoCoF ride-through capabilities are expected to stabilize the grid during disturbances by avoiding unnecessary disconnections that exacerbate grid instability.

\end{itemize}
We also look at the dispatch statistics of three individual facilities: Facility 13, Facility 2, and Facility 26 (highlighted in Table~\ref{tab:facilities}), examining the statistics in terms of Regulation (Raise/Lower) and Contingency (Raise/Lower) services over a year. The results are shown in Fig.~\ref{fig:kwinana}-\ref{fig:pinjar} in Appendix~\ref{apx:facilities}. Particularly, we observe that Facility 13, the only battery facility, is not used typically for Contingency Raise service. But gas-turbine generators are more frequently used for Contingency services. 

We list the outage records from \cite{aemo_realtime_outages} in Table~\ref{tab:facilities}, with the total outage duration (in hours) for ESS facilities over the past year. Surprisingly, many facilities recorded over 1K hours of outages, with one facility's unplanned outages reaching 10K hours. Although, as defined in \cite{aemo_market_data_wa}, an outage does not necessarily mean that the entire facility is out of service and there may still be some Remaining Available Capacity (RAC) of the facility during this time, the high frequency of these outages may be exploited by attackers. Especially for planned outages, which are typically announced in advance, this information could be useful for planned attacks.

\vspace{-5pt}
\subsection{\mbox{Statistics of Residential DPV \& Smart Inverters}}

To provide the context for our assessment study, we review the statistics of residential DPV penetration and smart inverters in Australia. Australia is characterized by high penetration of inverter-based resources with 1.2 million households. The uptake of small-scale solar or  DPV is expected to continue with about 34.55 GW of capacity to be installed in Australia by 2033-34. Currently, small-scale solar installations continue to be popular in Australia with around aggregate 40\% of households installed with rooftop DPV (nearing 18 GW of installed capacity). 

\smallskip
\noindent
$\triangleright$ {\bf Residential DPV Penetration:} Australia has a very high level of residential DPV installations. Based on data from \cite{apvi_pv_map} and \cite{energyBU_report}, the residential DPV capacity in Australia is projected to grow from 18.79 GW in 2024 to 28.32 GW by 2030, and further to 36.19 GW by 2035. Meanwhile, the DPV density\footnote{The {\em DPV density} refers to the ratio of the number of residential DPV installations over the number of dwellings in a given region.} in Australia is expected to increase from 45.04\% in 2023 to 69.16\% by 2030, reaching 86.51\% by 2035. We plotted the distribution of residential DPV density at each postcode in Australia for the years 2024 and the projection in 2035 in Fig.~\ref{fig:wa-pv-density}. It is evident that there is a significant upward trend in overall density, especially in the coastal regions, where many areas will reach 100\% density by 2030. The DPV density percentages across Australian states will significantly increase by 2030, with the overall DPV density in Australia projected to rise by 21.24\%. Note that DPV generation currently accounts for over 50\% of peak demand in most states. By 2030, DPV generation will exceed demand during peak periods in all but one of the states.

\smallskip
\noindent
$\triangleright$ {\bf Lifespan of Smart Inverters:}
Smart inverters are expected to have a relatively long lifecycle. We project the lifespan distribution of smart inverters in Australia in Fig.~\ref{fig:lifespan}, with the share of devices older than 15 years projected to grow from 0\% in 2015 to 40\% by 2045. We adopted the degradation rates from \cite{degradation}, where PV systems degrade by 0.7\% annually for the first 10 years, 0.46\% for the next 10 years, and 0.43\% thereafter. This aging trend poses significant cybersecurity risks, as older inverters are more susceptible to firmware vulnerabilities and may no longer receive regular updates. As the proportion of these vulnerable devices increases, they could become a convenient target for cyberattacks, potentially jeopardizing the security and reliability of the energy network.

\smallskip
\noindent
$\triangleright$ {\bf Market Share of Smart Inverters:}
We also provide insights into the market share of smart inverters in Australia. According to the report \cite{sunwiz2023inverter}, the 2022 market share of smart inverter brands in Australia is depicted in Fig.~\ref{fig:marketshare}. The top five brands are Sungrow (23\%), GoodWe (19\%), Fronius (19\%), Growatt (9\%), and SolarEdge (9\%). Together, these brands account for a significant portion of the total market share. Sungrow leads the market with a 23\% share. Brands from the same country are represented using similar color schemes. For instance,  Sungrow, GoodWe, Growatt, Ginlong/Solis and Huawei are from China. Fronius is from Austria, Enphase Energy is from the US, SolarEdge is from Israel, and SMA is from Germany. The predominance of smart inverters from one single country will exacerbate the risks of supply chain attacks.

\section{Data Analysis} \label{sec:analysis}

With the extensive open market datasets available from the Australia Electricity Market, we conducted a comprehensive data analysis for assessing the plausibility and impacts of power grid instability caused by cyberattacks on smart inverters. We specifically utilized the market data of Australia's frequency co-optimized Essential System Services (ESS) and dispatch schedules, along with other environmental data (e.g., solar irradiance), for an unnamed Australian state over the period of Oct 2023 - Sept 2024.

Our data analysis aims to address the following questions:
\begin{itemize}[leftmargin=*]

\item \underline{Sec~\ref{sec:analysis-opp}:} {\em What are the opportunities for wide-scale power instability triggered by concerted cyberattacks on smart inverters}? Will savvy attackers be able to glean insights to orchestrate attacks by harnessing data analysis from open dataset?

\item \underline{Sec~\ref{sec:analysis-freq}:} {\em What will be the possible impacts to the grid with respect to frequency control}? How likely can an attack cause system-wide instability, such as forced load shedding? 

\item \underline{Sec~\ref{sec:analysis-pred}:} {\em Can attackers maximize the impacts by predictive models}? Can machine learning techniques predict the optimal attack opportunities from open market data and facility data?

\end{itemize}

\vspace{-5pt}
\subsection{Assessment of Opportunities} \label{sec:analysis-opp}

\begin{figure}[b!]
\centering \vspace{-5pt}
\includegraphics[scale=0.25]{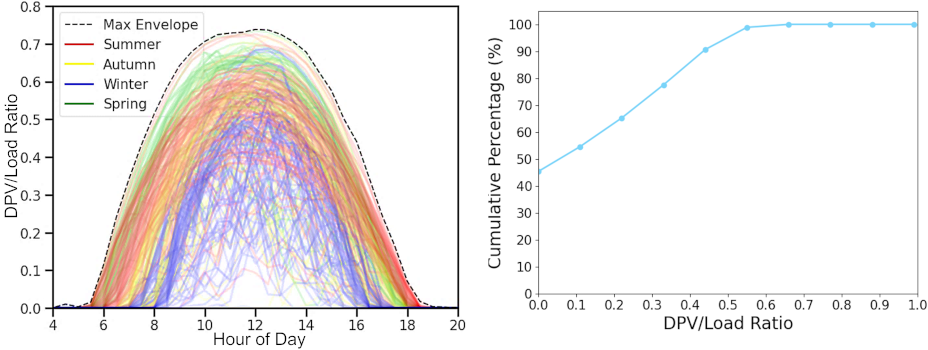}  \vspace{-10pt}
\caption{The ratios of distributed PV generation (DPV) over total load during a day in Oct 2023 - Sept 2024.} 
\label{fig:dpv-demand}
\end{figure}

\aptLtoX[graphic=no,type=html]{
\begin{figure}[t!]
\begin{minipage}{0.53\textwidth}
\centering
\includegraphics[width=1\textwidth]{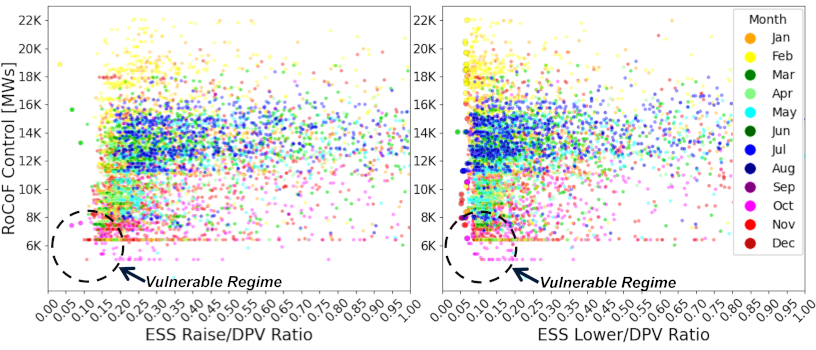}  \vspace{-20pt}
\caption{Joint distributions of ESS Raise or Lower to DPV ratios and RoCoF control for the occasions throughout a year.}  \vspace{-10pt}
\label{fig:dpv-ESS}
\end{minipage} 
\end{figure}
\begin{figure}[t!]
\begin{minipage}{0.46\textwidth}
\centering
\includegraphics[width=1.1\textwidth]{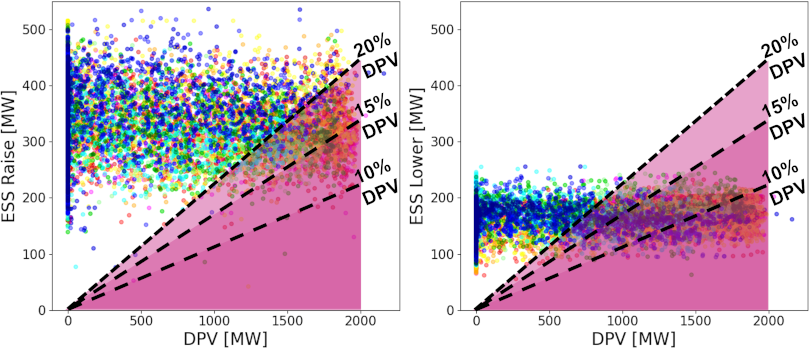}  \vspace{-20pt}
\caption{Joint distributions of ESS Raise/Lower vs DPV for the occasions throughout a year. }  \vspace{-10pt}
\label{fig:dpv-ESS2}
\end{minipage}
\end{figure}}{
\begin{figure*}[t!]
\begin{minipage}{0.53\textwidth}
\centering
\includegraphics[width=1\textwidth]{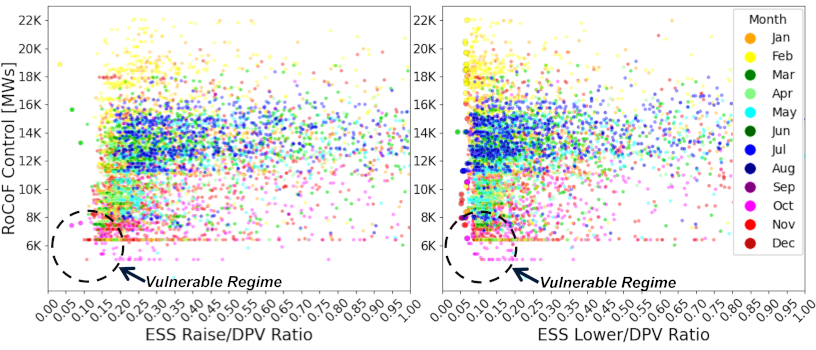}  \vspace{-20pt}
\caption{Joint distributions of ESS Raise or Lower to DPV ratios and RoCoF control for the occasions throughout a year.}  \vspace{-10pt}
\label{fig:dpv-ESS}
\end{minipage} 
\begin{minipage}{0.46\textwidth}
\centering
\includegraphics[width=1.1\textwidth]{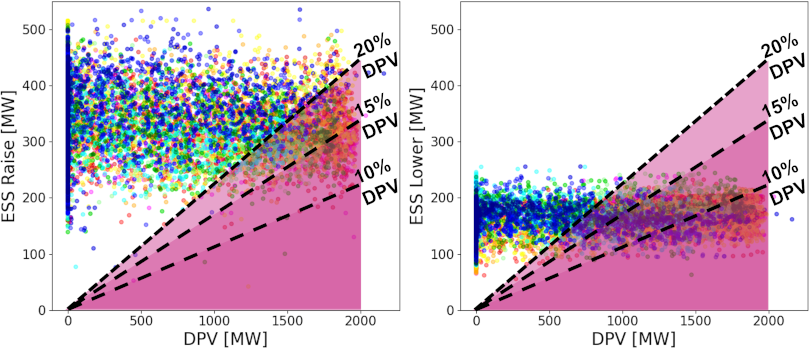}  \vspace{-20pt}
\caption{Joint distributions of ESS Raise/Lower vs DPV for the occasions throughout a year. }  \vspace{-10pt}
\label{fig:dpv-ESS2}
\end{minipage}
\end{figure*}}

\begin{figure}[t!]
\centering
\includegraphics[scale=0.25]{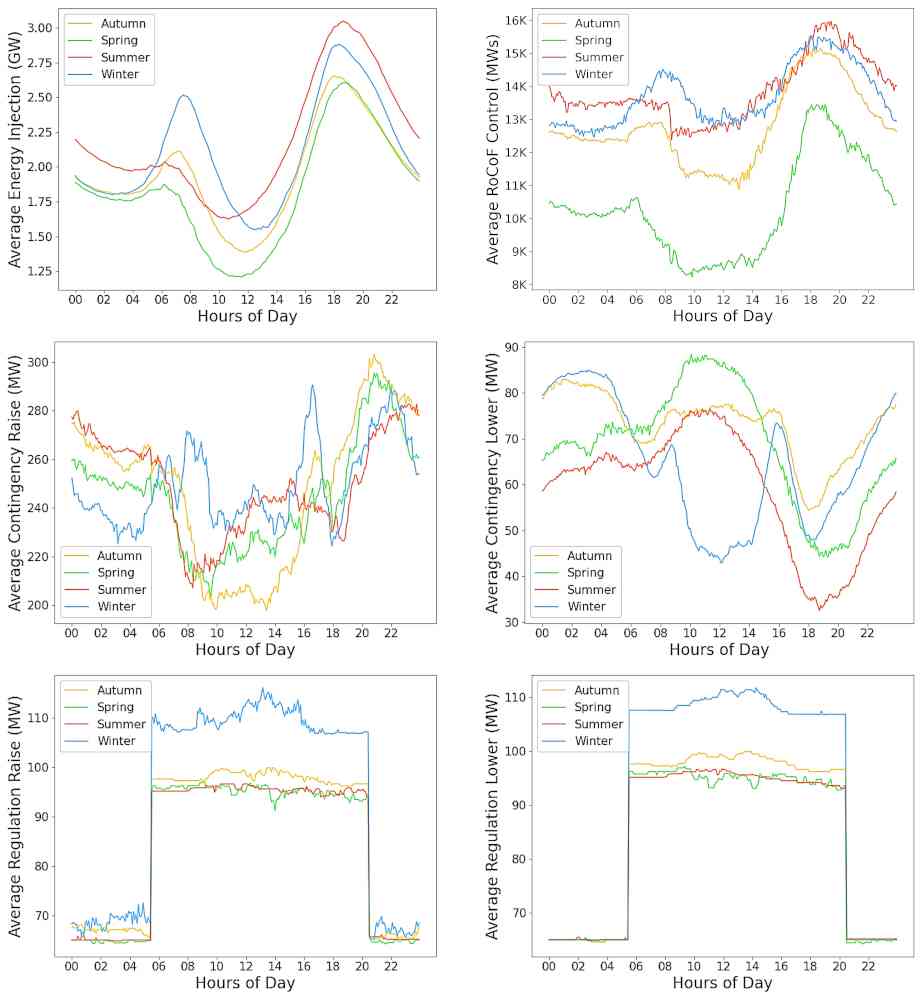}  \vspace{-10pt}
\caption{The average quantities of Energy Injection, RoCoF control, Contingency Raise/Lower and Regulation Raise/Lower during a day of each season.} \vspace{-15pt}
\label{fig:wem-stats}
\end{figure}

First, we define the following:
\begin{align}
\text{ESS Raise/DPV} & = \frac{\text{Contingency Raise} + \text{Regulation Raise}}{\text{DPV Generation}}
\label{eq:ess_raise_dpv} \\
\text{ESS Lower/DPV} & = \frac{\text{Contingency Lower} + \text{Regulation Lower}}{\text{DPV Generation}}
\label{eq:ess_lower_dpv}
\end{align}
Optimal opportunities can be characterized by two factors:
\begin{enumerate}[leftmargin=*]

\item {\bf\em Low RoCoF control} quantity would be beneficiary to attackers, as there will be less system inertia in the grid to counteract the changes of frequency (see more details in Sec.~\ref{sec:analysis-freq}). 

\item {\bf\em Low ESS Raise/DPV ratio or low ESS Lower/DPV ratio} also presents a good attack opportunity, as the attackers would need a smaller portion of DPV to exceed the limits of what ESS Raise/Lower services can compensate. 

\end{enumerate}

We plot the ratios of DPV generation over total load (i.e. DPV/load) during a day in Fig.~\ref{fig:dpv-demand}. We observe that the maximum ratios were attained in spring, and the peak ratio is about 75\%  at noon time. Also, the distribution of DPV/load ratios is moderate near the peak ratio. This signifies ample opportunities of high DPV generation.

Next, we plot the average quantities of Energy Injection for energy adequacy, RoCoF control, Contingency Raise/Lower and Regulation Raise/Lower services during a day of each season in Fig.~\ref{fig:wem-stats}. We observe the following patterns:

\begin{enumerate}[leftmargin=*]

     \item {\em Energy Injection/RoCoF Control:} 
     Both quantities display clear diurnal patterns, with peaks in the morning and afternoon, and dips around midday. There is a strong correlation between the RoCoF control and Energy Injection quantities, as their trends mirror each other throughout the day.
     
    \item {\em Contingency Raise/Lower:}  
    The Contingency Raise quantity tends to be higher during the early and late hours, with noticeable fluctuations and two peaks at midnight and midday. Contingency Lower quantity, on the other hand, generally exhibits an opposite trend to the Contingency Raise quantity and shows a positive correlation with Energy Injection and RoCoF quantities. Spring and summer follow similar patterns, while winter displays a distinct pattern.

    \item {\em Regulation Raise/Lower:}  
    Both quantities exhibit identical patterns, with low capacity between midnight and 6am, higher capacity from 6am to 8pm, and a decrease after 8 pm. Winter consistently shows the highest Regulation Raise, while the other seasons display less pronounced differences. The variation in Regulation Lower closely mirrors that of Regulation Raise.

\end{enumerate}

\noindent
{\bf\em Ramifications}: Because of these patterns, we conclude that (1) ESS Raise/DPV ratio is typically low during the noon time when DPV generation is high; (2) ESS Lower/DPV ratio is especially low in winter during noon time; (3) RoCoF control quantity is also low during noon time. Hence, there is a clear misalignment of contingency capacity with DPV, which is due to the fact that the existing contingency mechanisms typically focus on inadvertent events like the failure of a fossil-fuelled generator, without considering the possibilities of cyberattacks on DPV.

Moreover, we plot the joint distributions of ESS Raise or Lower to DPV ratios and RoCoF control for all the occasions throughout a year in Fig.~\ref{fig:dpv-ESS}. We particularly highlight the {\em vulnerable regimes} that are characterized by low ESS Raise/DPV ratio and low RoCoF control, as well as low ESS Lower/DPV ratio and low RoCoF control. We note that the occasions in the vulnerable regime have an ESS Raise/DPV ratio as low as 10-15\%, and an ESS Lower/DPV ratio as low as 5-10\%, which implies a low barrier for attacks. Fig.~\ref{fig:dpv-ESS2} shows that a considerable portion of ESS Raise and Lower quantities are lying beneath 10-20\% of DPV.

\aptLtoX[graphic=no,type=html]{\begin{figure}[t!]
\begin{minipage}{0.46\textwidth}
\centering 
\scriptsize
\resizebox{0.75\columnwidth}{!}{
\begin{tabular}{@{}ccccccc@{}}
Date & \rot{Load (GW)} & \rot{DPV (GW)} & \rot{RoCoF (MWs)} & \rot{ESS Raise (GW)}  & \rot{DPV Loss (GW)}  & \rot{Time to UFLS (sec)} \\
\hline
09 Dec 23 11:00am    & 2.67 & 1.93 & 6.4K & 0.22 & 0.24 & 4.42 \\
09 Dec 23 11:00am    & 2.61 & 1.89 & 6.4K & 0.22 & 0.24 & 4.43 \\
10 Dec 23 11:30am & 2.66 & 1.94 & 6.4K & 0.22 & 0.24 & 4.43 \\
09 Dec 23 12:00pm    & 2.69 & 1.94 & 6.4K & 0.22 & 0.24 & 4.43 \\
09 Dec 23 12:30pm & 2.68 & 1.93 & 6.4K & 0.22 & 0.24 & 4.43 \\
09 Dec 23 01:30pm  & 2.63 & 1.79 & 6.4K & 0.21 & 0.23 & 4.48 \\
09 Dec 23 01:00pm     & 2.65 & 1.87 & 6.4K & 0.21 & 0.23 & 4.46 \\
09 Dec 23 02:00pm     & 2.57 & 1.69 & 6.4K & 0.20 & 0.22 & 4.51 \\
10 Dec 23 09:30am  & 2.42 & 1.58 & 6.4K & 0.18 & 0.20 & 5.53 \\
09 Dec 23 10:00am    & 2.51 & 1.71 & 6.4K & 0.18 & 0.20 & 5.52 \\
09 Dec 23 10:30am & 2.58 & 1.81 & 6.4K & 0.18 & 0.20 & 5.53 \\
17 Oct 23 11:00am    & 3.23 & 1.86 & 7.6K & 0.22 & 0.24 & 8.52 \\
10 Nov 23 01:00pm     & 3.14 & 1.86 & 5K & 0.20 & 0.22 & 7.40 \\
\hline
\end{tabular}
}\vspace{-5pt}
\caption{Possible DPV loss attack scenarios in the vulnerable regime in Fig.~\ref{fig:dpv-ESS} (Left).}
\label{tab:DPV_loss}
\end{minipage} \quad
\end{figure}
\begin{figure}
\begin{minipage}{0.47\textwidth}
\centering 
\scriptsize
\resizebox{0.85\columnwidth}{!}{
\begin{tabular}{@{}ccccccc@{}}
Date & \rot{Load (GW)} & \rot{DPV (GW)} & \rot{RoCoF (MWs)} & \rot{ESS Lower (GW)}  & \rot{DPV Hike (GW)}  & \rot{Time to OFGS (sec)} \\
\hline
10 Nov 23 01:00pm  & 3.14 & 1.86 & 5K & 0.20 & 0.22 & 5.69 \\
30 Oct 23 02:30pm & 2.57 & 1.48 & 5K & 0.17 & 0.19 & 5.69 \\
30 Oct 23 02:00pm & 2.65 & 1.62 & 5.1K & 0.17 & 0.19 & 5.69 \\
30 Oct 23 01:30pm & 2.67 & 1.75 & 5.4K & 0.17 & 0.19 & 5.70 \\
30 Oct 23 01:00pm & 2.79 & 1.85 & 5.4K & 0.17 & 0.19 & 5.70 \\
30 Oct 23 11:00am & 2.85 & 1.91 & 5.4K & 0.17 & 0.19 & 5.70 \\
30 Oct 23 11:30am & 2.86 & 1.94 & 5.4K & 0.17 & 0.19 & 5.70 \\
30 Oct 23 12:00pm & 2.85 & 1.95 & 5.4K & 0.17 & 0.19 & 5.71 \\
30 Oct 23 12:30pm & 2.85 & 1.91 & 5.4K & 0.17 & 0.19 & 5.71 \\
30 Oct 23 09:30am & 2.70 & 1.60 & 6.1K & 0.17 & 0.19 & 5.71 \\
11 Oct 23 09:30am & 2.64 & 1.46 & 5.8K & 0.17 & 0.19 & 5.71 \\
30 Oct 23 10:30am & 2.74 & 1.83 & 5.4K & 0.17 & 0.19 & 5.70 \\
28 Oct 23 10:00am & 2.52 & 1.74 & 6.1K & 0.14 & 0.15 & 7.74 \\
\hline
\end{tabular}
}\vspace{-5pt}
\caption{Possible DPV hike attack scenarios in the vulnerable regime in Fig.~\ref{fig:dpv-ESS} (Right).}
\label{tab:DPV_hike}
\end{minipage} \vspace{-5pt}
\end{figure}
}{
\begin{figure*}[t!]
\begin{minipage}{0.47\textwidth}
\centering \vspace{15pt}
\scriptsize
\resizebox{0.85\columnwidth}{!}{
\begin{tabular}{@{}ccccccc@{}}
Date & \rot{Load (GW)} & \rot{DPV (GW)} & \rot{RoCoF (MWs)} & \rot{ESS Raise (GW)}  & \rot{DPV Loss (GW)}  & \rot{Time to UFLS (sec)} \\
\hline
09 Dec 23 11:00am    & 2.67 & 1.93 & 6.4K & 0.22 & 0.24 & 4.42 \\
09 Dec 23 11:00am    & 2.61 & 1.89 & 6.4K & 0.22 & 0.24 & 4.43 \\
10 Dec 23 11:30am & 2.66 & 1.94 & 6.4K & 0.22 & 0.24 & 4.43 \\
09 Dec 23 12:00pm    & 2.69 & 1.94 & 6.4K & 0.22 & 0.24 & 4.43 \\
09 Dec 23 12:30pm & 2.68 & 1.93 & 6.4K & 0.22 & 0.24 & 4.43 \\
09 Dec 23 01:30pm  & 2.63 & 1.79 & 6.4K & 0.21 & 0.23 & 4.48 \\
09 Dec 23 01:00pm     & 2.65 & 1.87 & 6.4K & 0.21 & 0.23 & 4.46 \\
09 Dec 23 02:00pm     & 2.57 & 1.69 & 6.4K & 0.20 & 0.22 & 4.51 \\
10 Dec 23 09:30am  & 2.42 & 1.58 & 6.4K & 0.18 & 0.20 & 5.53 \\
09 Dec 23 10:00am    & 2.51 & 1.71 & 6.4K & 0.18 & 0.20 & 5.52 \\
09 Dec 23 10:30am & 2.58 & 1.81 & 6.4K & 0.18 & 0.20 & 5.53 \\
17 Oct 23 11:00am    & 3.23 & 1.86 & 7.6K & 0.22 & 0.24 & 8.52 \\
10 Nov 23 01:00pm     & 3.14 & 1.86 & 5K & 0.20 & 0.22 & 7.40 \\
\hline
\end{tabular}
}\vspace{-5pt}
\caption{Possible DPV loss attack scenarios in the vulnerable regime in Fig.~\ref{fig:dpv-ESS} (Left).}
\label{tab:DPV_loss}
\end{minipage} \quad
\begin{minipage}{0.47\textwidth}
\centering \vspace{15pt}
\scriptsize
\resizebox{0.85\columnwidth}{!}{
\begin{tabular}{@{}ccccccc@{}}
Date & \rot{Load (GW)} & \rot{DPV (GW)} & \rot{RoCoF (MWs)} & \rot{ESS Lower (GW)}  & \rot{DPV Hike (GW)}  & \rot{Time to OFGS (sec)} \\
\hline
10 Nov 23 01:00pm  & 3.14 & 1.86 & 5K & 0.20 & 0.22 & 5.69 \\
30 Oct 23 02:30pm & 2.57 & 1.48 & 5K & 0.17 & 0.19 & 5.69 \\
30 Oct 23 02:00pm & 2.65 & 1.62 & 5.1K & 0.17 & 0.19 & 5.69 \\
30 Oct 23 01:30pm & 2.67 & 1.75 & 5.4K & 0.17 & 0.19 & 5.70 \\
30 Oct 23 01:00pm & 2.79 & 1.85 & 5.4K & 0.17 & 0.19 & 5.70 \\
30 Oct 23 11:00am & 2.85 & 1.91 & 5.4K & 0.17 & 0.19 & 5.70 \\
30 Oct 23 11:30am & 2.86 & 1.94 & 5.4K & 0.17 & 0.19 & 5.70 \\
30 Oct 23 12:00pm & 2.85 & 1.95 & 5.4K & 0.17 & 0.19 & 5.71 \\
30 Oct 23 12:30pm & 2.85 & 1.91 & 5.4K & 0.17 & 0.19 & 5.71 \\
30 Oct 23 09:30am & 2.70 & 1.60 & 6.1K & 0.17 & 0.19 & 5.71 \\
11 Oct 23 09:30am & 2.64 & 1.46 & 5.8K & 0.17 & 0.19 & 5.71 \\
30 Oct 23 10:30am & 2.74 & 1.83 & 5.4K & 0.17 & 0.19 & 5.70 \\
28 Oct 23 10:00am & 2.52 & 1.74 & 6.1K & 0.14 & 0.15 & 7.74 \\
\hline
\end{tabular}
}\vspace{-5pt}
\caption{Possible DPV hike attack scenarios in the vulnerable regime in Fig.~\ref{fig:dpv-ESS} (Right).}
\label{tab:DPV_hike}
\end{minipage} 
\end{figure*}
}

\begin{figure}[h!]
\centering
\scriptsize
\includegraphics[scale=0.25]{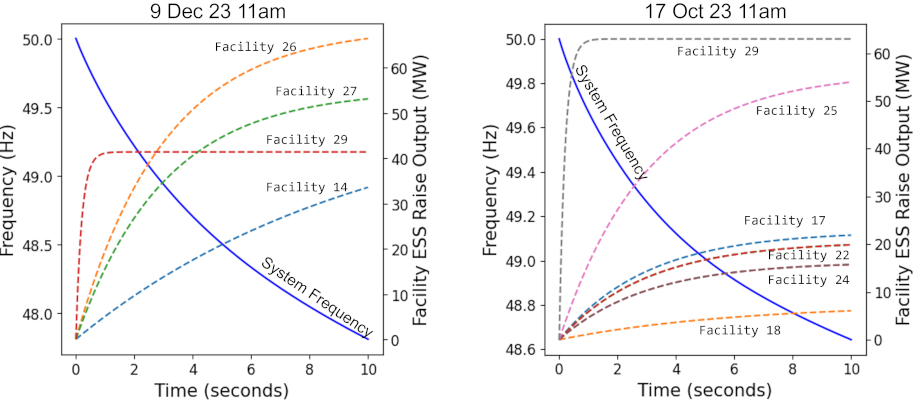}  \vspace{-5pt}
\caption{Emulating DPV loss attacks on 9 Dec \& 17 Oct 23.} \vspace{-5pt}
\label{fig:freq_essraise}
\end{figure}

\vspace{-5pt}
\subsection{Impacts on Frequency Control} \label{sec:analysis-freq}

In this section, we emulate attacks on frequency control with the occasions in the vulnerable regimes in Fig.~\ref{fig:dpv-ESS}. To assess the impacts on frequency control, we model the RoCoF and its relation to the inertial in the grid based on the {\em single-mass machine theory}. The basic idea is that the whole power grid can be treated as a single machine with aggregated kinetic energy (and inertia) of all activated generators and a uniform frequency is assumed throughout the grid. The single-mass machine theory is often justified by examining empirical grid disturbances and the frequencies throughout the grid measured using GPS time-synchronized sensors.

\smallskip
\noindent
$\triangleright$ {\bf RoCoF Model:}
By the single-mass machine theory, the {\em rate of change of frequency} (RoCoF) in the grid with respect to the imbalance between generation and load, and system inertia is modelled by \cite{AM90inertia}: \vspace{-5pt}
\begin{equation}
\frac{d f(t)}{dt} = \frac{f_n}{2}\cdot\frac{\Delta P(t)}{{\sf K_{sys}}} 
\label{eqn:derivative}
\end{equation}
where $f(t)$ is the frequency of the grid at time $t$, $f_n$ is the initial frequency (50Hz as the normal operating frequency), $\Delta P(t)$ is the imbalance between generation and load (MW) at time $t$, and ${\sf K_{sys}}$ is total system inertia (MWs).

Note that ${\sf K_{sys}}$ includes the synchronous inertia from generators ${\sf K_{gen}}$ and load ${\sf K_{load}}$. Generator inertia ${\sf K_{gen}}$ can be obtained from the market data on RoCoF control, whereas load inertia ${\sf K_{load}}$ can be inferred by the load. The study in \cite{aemo21inertia} presents an empirical estimation of the load inertia in Australia's power grid based on the observed load. We adopt the regression model from their study to estimate the load inertia ${\sf K_{load}}$ for our data analysis.

When there is an imbalance between generation and load, some contingency facilities will be activated to compensate for the imbalance. However, these facilities cannot be increased to their full capacity instantly. The temporal response of a facility's generation can be modelled by the following equation: %\vspace{-5pt}
\begin{equation} 
P(t) = P_{\sf max} (1 - e^{-\frac{t}{\tau}})
\label{eqn:response}
\end{equation}
where $P(t)$ is the generation at time $t$, $P_{\sf max}$ is the maximum capacity and $\tau$ is the speed factor that characterizes the responsiveness of a facility. For instance, battery storage will have a faster response time (i.e., smaller speed factor) than synchronous generators, like coal-fired generators. Note that Australia's electricity markets published the speed factors of all contingency facilities.  

When there is a shortage of DPV generation at time $t_0$, $\Delta P(t_0)$ and $\frac{d f(t)}{dt}$ will be at the maximum. As more contingency facilities are increasing the generation level to compensate for the imbalance, $\Delta P(t_0)$ will decrease and so will RoCoF.

\begin{figure*}[t!]
    \centering \vspace{-10pt}
    \subfigure[Actual vs predicted ESS Raise DPV Ratio.]{
        \includegraphics[width=0.32\linewidth]{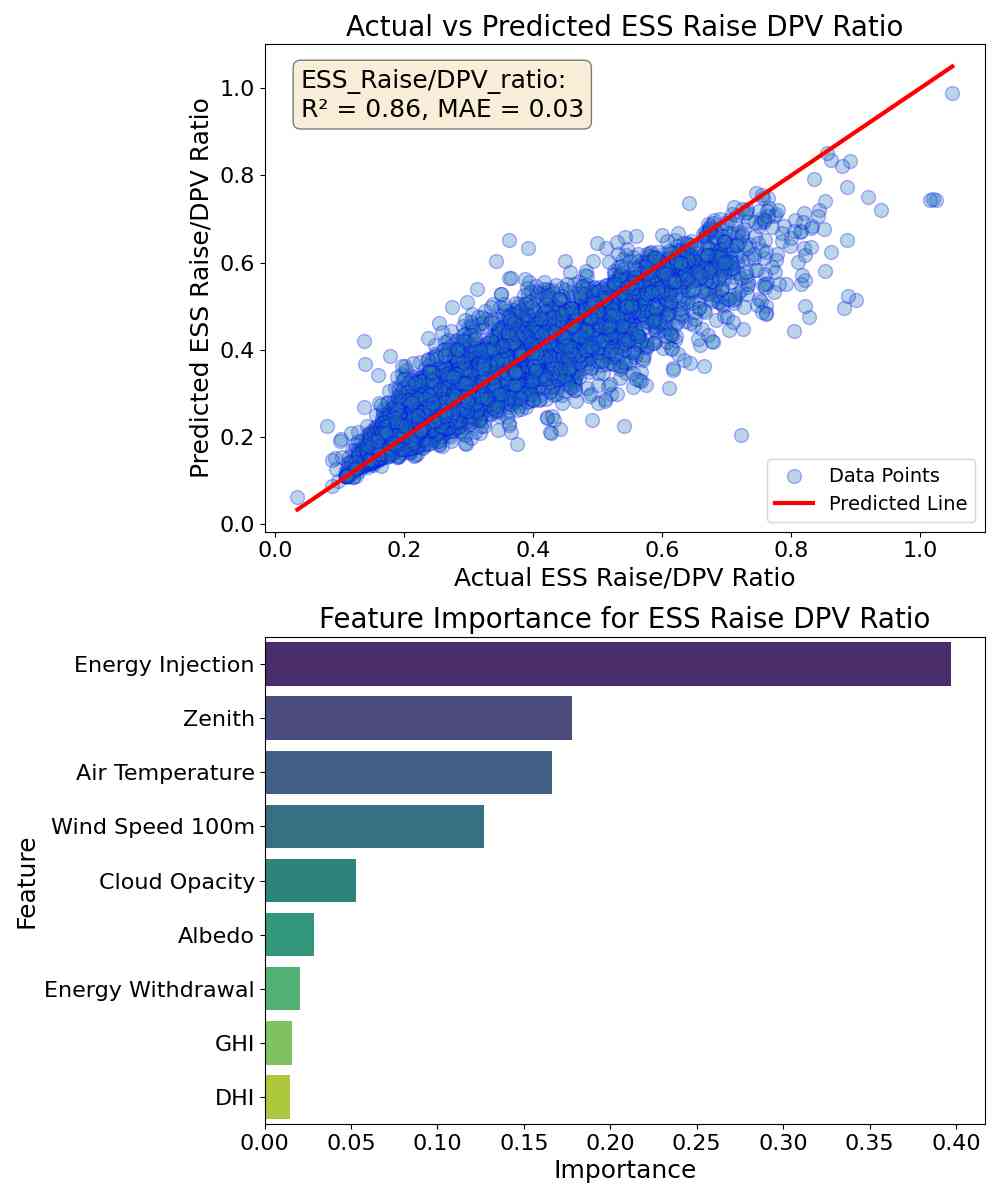}
        \label{fig:ess_raise_vs_feature_importance}
    }
    \hfill
    \subfigure[Actual vs predicted ESS Lower DPV Ratio.]{
        \includegraphics[width=0.32\linewidth]{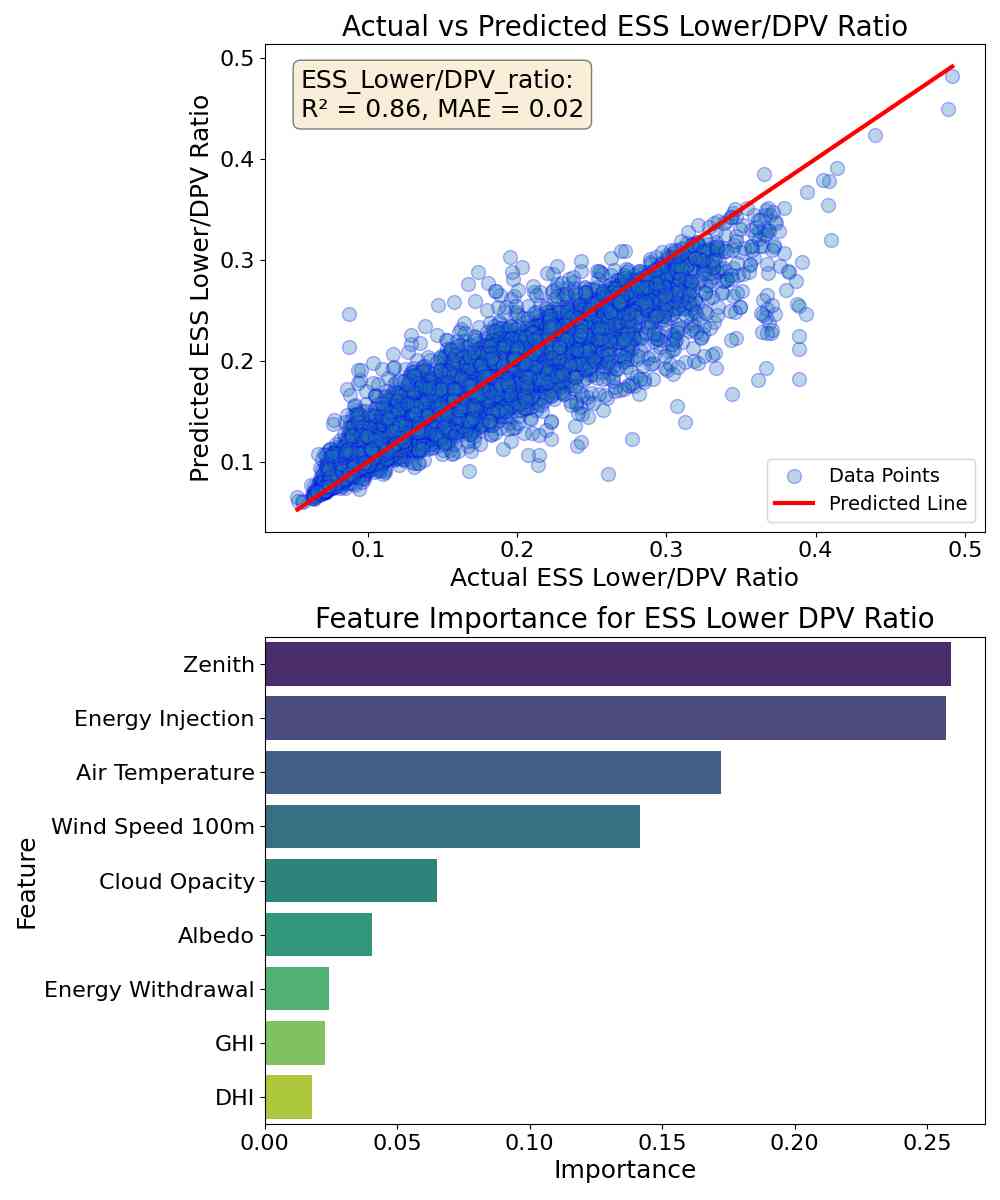}
        \label{fig:ess_lower_vs_feature_importance}
    }
    \hfill
    \subfigure[Actual vs predicted RoCoF values.]{
        \includegraphics[width=0.32\linewidth]{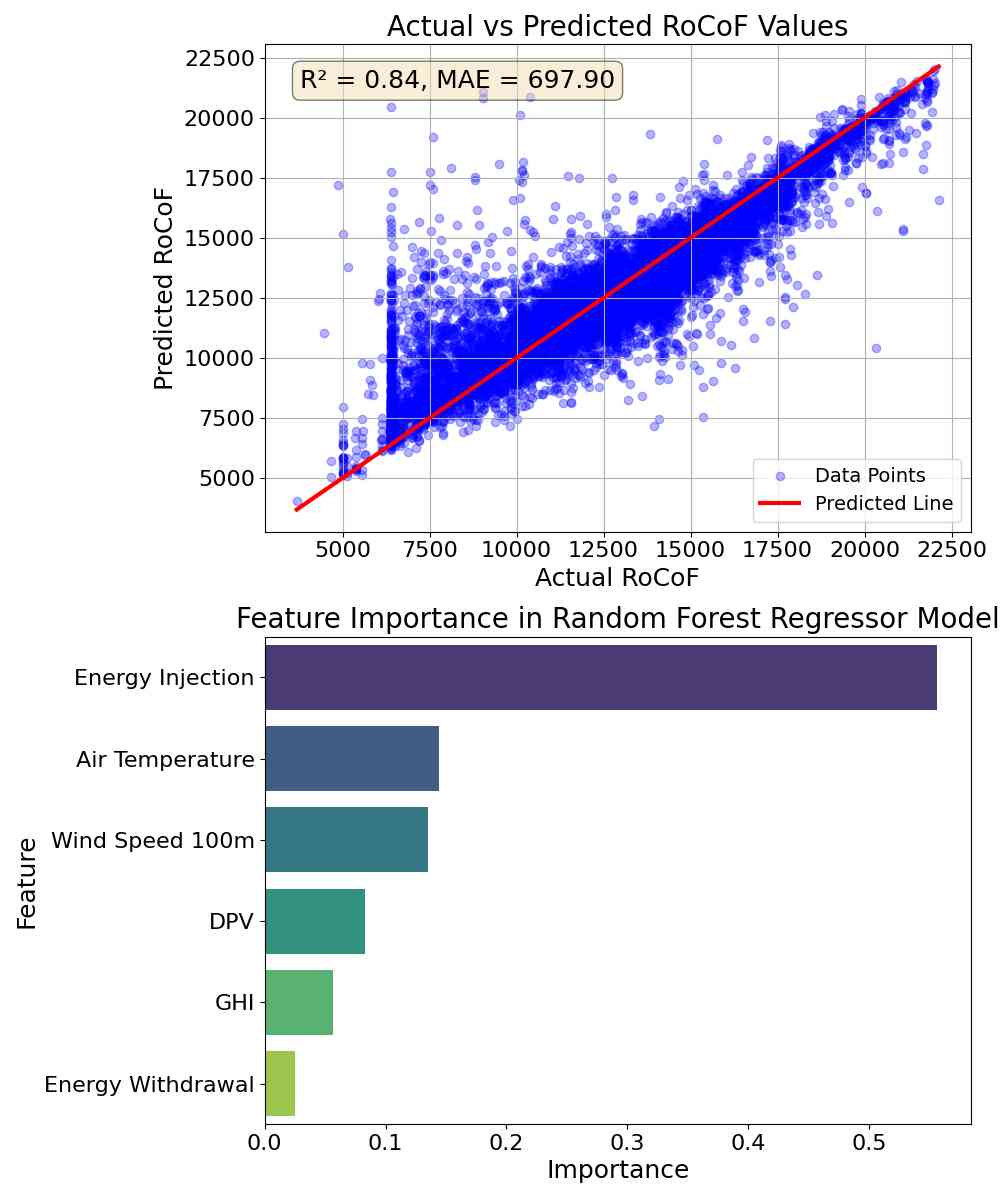}
        \label{fig:predicted_RoCoF}
    }
    \vspace{-10pt}
    \caption{Predictions of ESS Raise, ESS Lower, and RoCoF, along with a comparison of feature importance.}
    \label{fig:combined_figures} \vspace{-5pt}
\end{figure*}

\smallskip
\noindent
$\triangleright$ {\bf Attacks on Frequency Control:} We consider two types of attacks on frequency control by compromised smart inverters: 
\begin{itemize}[leftmargin=*]

\item {\bf\em DPV Loss Attack}: An attacker turns off the compromised smart inverters to cause a loss of DPV generation. The threshold to trigger UFLS (Under Frequency Load Shedding) is 48.75Hz \cite{aemo_freqresp}.

\item {\bf\em DPV Hike Attack}: An attacker initially turns off the compromised smart inverters and then waits for an opportunity to suddenly turn on the inverters to cause a hike of DPV generation. The threshold to trigger OFGS (Over Frequency Generation Shedding) is 52Hz \cite{aemo_freqresp}.

\end{itemize}
We measure the time for the frequency $f(t)$ to hit the threshold for UFLS or OFGS when there is a continual imbalance in the grid. We emulate the RoCoF under attacks based on Eqns~(\ref{eqn:derivative})-(\ref{eqn:response}).

 In Table~\ref{tab:DPV_loss}, we list the possible DPV loss attack scenarios in the vulnerable regime in Fig.~\ref{fig:dpv-ESS}. We simulated the system response, where the DPV loss is 110\% of ESS Raise, with an imbalance of 10\% of ESS Raise. The ``Time to UFLS'' indicates the time it would take for the system frequency to drop to 48.75 Hz, triggering UFLS. In Fig.~\ref{fig:freq_essraise}, we especially depict the simulated frequency response of and output of the dispatched ESS Raise facilities under DPV loss attacks at 11 am on 9 Dec and 17 Oct 23.

 In Table~\ref{tab:DPV_hike}, we list the possible DPV loss attack scenarios in the vulnerable regimes in Fig.~\ref{fig:dpv-ESS}. We simulated the system response, where the DPV hike is 110\% of ESS Lower, with an imbalance of 10\% of ESS Lower. The ``Time to OFGS'' represents the time it would take for the system frequency to rise to 52 Hz, triggering OFGS.

\smallskip
\noindent
{\bf\em Ramifications}:
We observe that under both attack scenarios, small-scale attacks ($\sim$ 0.2 GW DPV) are sufficient to trigger UFLS or OFGS. Hence, attackers only need a small portion to launch an impactful attack to cause wide-scale grid instability.

\vspace{-5pt}
\subsection{Predictive Models for Opportunities}
\label{sec:analysis-pred}

In this section, we study how attackers can leverage machine learning techniques to predict the optimal attack opportunities from open market and facility data. Our predictive models show that the reliance on ESS for frequency control is strongly influenced by weather and market variables outlined in Appendix~\ref{apx:model_factors}. To infer attack opportunities under different conditions, we aim to predict the ESS Raise/DPV, ESS Lower/DPV ratios and RoCoF.

{\bf\em Model Overview:}  
We employed the {\tt RandomForest-Regressor} tool from {\tt sklearn.ensemble} in Python to analyze the relationship between grid dynamics and key influencing factors. This ensemble learning method aggregates predictions from multiple decision trees, effectively capturing non-linear relationships and interactions between variables. Its robustness against overfitting and ability to model intricate dependencies make it well-suited for predicting grid behaviors under diverse weather and market conditions. The models are defined as follows:
\[
Y_{\text{ESS}} \triangleq f(X_{\text{weather}}, X_{\text{market}}),\quad
Y_{\text{RoCoF}} \triangleq f(X_{\text{weather}}, X_{\text{market}}, X_{\text{system}})
\]
where \(Y_{\text{ESS}}\) represents the predicted ESS Raise/DPV or ESS Lower/DPV ratio, and \(Y_{\text{RoCoF}}\) represents the predicted Rate of Change of Frequency (RoCoF). 

{\bf\em Feature Sets and Input Variables:}  
The models utilize three sets of input features:
\begin{itemize}
    \item \(X_{\text{weather}}\): Environmental variables, including air temperature, zenith angle, cloud opacity, direct horizontal irradiance (DHI), and global horizontal irradiance (GHI).
    \item \(X_{\text{market}}\): Market-related variables, such as energy injection capacity and energy withdrawal capacity.
    \item \(X_{\text{system}}\): System-level variables, including DPV generation.
\end{itemize}

$\triangleright$ {\bf Data Preprocessing and Model Training:}  
To ensure high data quality and meaningful predictions, we filtered out instances where DPV is below a threshold of 500 MW. Such filtering addresses the higher noise levels typically observed during early mornings, late evenings, or extreme weather conditions when DPV generation is minimal. These conditions are characterized by rapid fluctuations and lower data reliability, which could otherwise degrade model accuracy. By focusing on intervals with substantial DPV generation, the models were better equipped to capture meaningful patterns and reduce the influence of noisy data.

The dataset, comprising a total of 30,270 data points, was then randomly split into training and testing sets in a 7:3 ratio, resulting in 21,189 data points in the training set and 9,081 data points in the testing set. This approach helps to prevent overfitting and provides a robust assessment of the model's generalization capabilities.

{\bf\em Model Performance and Feature Importance Analysis:}
Fig.~\ref{fig:ess_raise_vs_feature_importance} shows the model performance for the prediction of the ESS Raise/DPV ratio with very high accuracy. The actual versus predicted plot on the left shows a strong alignment. The feature importance analysis on the right shows that energy injection capacity and zenith angle emerge as the most influential variables, emphasizing the critical roles these variables play in maintaining grid stability. Energy injection capacity directly affects the amount of power that can be added to the grid during frequency disturbances, while the zenith angle is related to solar irradiance.

Similarly, Fig.~\ref{fig:ess_lower_vs_feature_importance} shows that the predictive model for the ESS Lower/DPV ratio also performs with high accuracy. The alignment between actual and predicted values demonstrates the model's strength in predicting when the grid will rely on ESS to manage excess energy. The feature importance analysis identifies the zenith angle and energy injection capacity as the most critical variables, significantly impacting power grid behavior and solar irradiance, which in turn influences DPV generation.

Fig.~\ref{fig:predicted_RoCoF} depicts the final model representing the grid's capability to handle changes in frequency due to RoCoF. Various predicted versus actual RoCoF values depicted on the left demonstrate high correlation values, proving how well this model captures the dynamic response of the grid to generation-load imbalance conditions. The feature importance analysis on the right shows that energy injection capacity is the most significant feature influencing RoCoF, followed by air temperature and wind speed. This underlines the importance of grid capacity for injecting energy in case of frequency disturbances and how this is interrelated with environmental factors for frequency stability.

\medskip
\noindent
{\bf\em Ramifications}:
By combining these observations together, one can see how interdependent many components of ESS are. The ESS Raise and Lower ratios provide an indication of the current reliance of the grid on the frequency control services in times of stressors such as extreme environmental change, while RoCoF gives insight into the capacity of the grid to resist rapid frequency changes. These models offer insights on the predictability of attack opportunities, and hence reinforce that attackers can maximize the impacts by advanced leveraging machine learning to predict the optimal attack opportunities. Note that there is further open facility data (e.g., planned/unplanned outages), which can be utilised to improve the accuracy of predicting optimal attack opportunities.

%\vspace{-5pt}
\section{Discussion and Conclusion} \label{sec:discuss}

In this paper, we showed that the threat of cyberattacks on smart inverters causing wide-scale power instability is significant. Particularly, the contingency capacity mechanisms for power grids are conventionally designed to cope with inadvertent contingency events, which are insufficient to fend off savvy attackers. 

\medskip

We conclude this paper with a discussion of further ramifications:

\smallskip

{\bf\em Goals of Attacks:}  
Cyberattacks on smart inverters can be motivated by various purposes. Besides terrorism, extortion and blackmailing, we note that attackers may be financially motivated to profit from the instability of energy market. For instance, the Australian Securities Exchange (ASX) Energy trades futures and options with respect to energy market prices \cite{asx}. Attackers, who short energy market prices using derivatives on ASX Energy, may profit from the fluctuations in energy market prices as a result of cyberattacks on energy infrastructure. It is worth mentioning that there was an unprecedented surge of the energy market spot price in Jun 2022, due to the suspension of Australia's National Electricity Market (NEM) after the withdrawal of a large volume of capacity \cite{aemo_suspend22}, which can be exploited by cyberattacks on power grid. 

{\bf\em Limitations of Findings:} While this study offers insights on the impacts of cyberattacks on smart inverters, the findings are subject to several limitations. First, the effectiveness of frequency control attacks relies on the actual level of load inertia. The level of load inertia is hard to estimate precisely. A possible error in the estimation would contribute to a considerable deviation of attack opportunities. Second, covert coordination of compromised DPV is a challenging task. Attackers may be detected during the coordination stage, which will thwart the launches of attacks. Third, the compromised DPV devices will be disconnected once an attack is detected, preventing them from future attacks. Hence, attackers may not be able launch regular attacks using compromised DPV devices, which limits the further impacts of their attacks. Despite the limitations, our study identifies general insights on the weaknesses of power grid against cyberattacks on smart inverters and improves the understanding of robust strategies against such attacks.

{\bf\em Mitigation Strategies:}  
The paper identifies a misalignment of contingency capacity with DPV generation that leads to attack opportunities with low DPV. Note that this is not the issue due to contingency capacity's correlation with the possibilities of inadvertent events (which are manifested in energy injection). But the limited correlation with only energy injection will present an opportunity for attackers. A natural mitigation strategy is to broaden the correlation of contingency capacity with DPV generation, which will significantly increase the barrier for attackers. Emergency solar management system \cite{ESM-AMEO} that is designed to curtail excessive DPV generation may also provide a responsive countermeasure to limit DPV hike attacks. Further assessment will be pursued to investigate the effectiveness of these mitigation strategies in future work.

{\bf\em Implications to Other Places:}  
Our study casts insights on defending the grid in the presence of cyberattacks, highlighting the need for Frequency Co-optimised Essential System Services (ESS) that explicitly consider prospective cyberattacks on consumer energy resources. We note that market-based co-optimized frequency control mechanisms are emerging worldwide. For example, California ISO (CAISO), Texas ISO (ERCOT), UK ISO and Ireland ISO adopted a similar ESS design in their markets \cite{SusantoReport}. UK ISO and Ireland ISO also adopted a similar system inertia control. But there is no known open historic dataset or assessment in these places. The lessons from our study can shed light on the robust ESS design for places with high penetration of DPV.

%\section{Conclusion} \label{sec:concl}

\begin{acks}
This project was funded by Cyber Security Cooperative Research Centre (C11-00306) and CSIRO's Critical Infrastructure Protection and Resilience Mission (R-20215).
\end{acks}

\clearpage
\bibliographystyle{ACM-Reference-Format}
\bibliography{references}

%\clearpage
\appendix
%\clearpage
\section{Appendix}

\subsection{Additional Information for ESS Facilities} \label{apx:facilities}

In the unnamed Australian state, while there are 82 facilities registered for Frequency Co-optimised Essential System Services (ESS), there are only 29 active facilities that provided ESS in the past year. In Table~\ref{tab:facilities}, we list the data of all active ESS facilities in that state \cite{aemo_fcess_accreditation}. The table includes details on the generator types and the associated maximum capacities in MW for different ESS classes. The table highlights the utilization rates of these facilities over the period of Oct 2023 - Sept 2024. 

We particularly look at the dispatch statistics of three individual facilities: Facility 13, Facility 2, and Facility 26 (highlighted in pink in Table~\ref{tab:facilities}).  We examine the statistics in terms of Regulation (Raise/Lower) and Contingency (Raise/Lower) services over a year. The statistics plots are depicted in Fig.~\ref{fig:kwinana}-\ref{fig:pinjar}.

From the dispatch statistics, we observe that Facility 13, a battery facility, is used for both Regulation and Contingency services. However, its most significant contribution is to Regulation Lower, where a clear pattern emerges: from 8 PM to 6 AM, the output remains stable at around 50 MW, while from 6 AM to 8 PM, the output most often fluctuates around 90 MW.
For Facility 2 (Gas), there is virtually no contribution to Raise services, with the majority of its output focused on Lower. On the other hand, Facility 26 (Gas) exhibits the opposite trend, with most of its output concentrated on Raise services.

% \clearpage
% \newpage

\subsection{Impact of Market and Environmental Factors on ESS Ratios} \label{apx:model_factors}

\begin{figure*}[h!]
    \centering
    \includegraphics[width=0.8\linewidth]{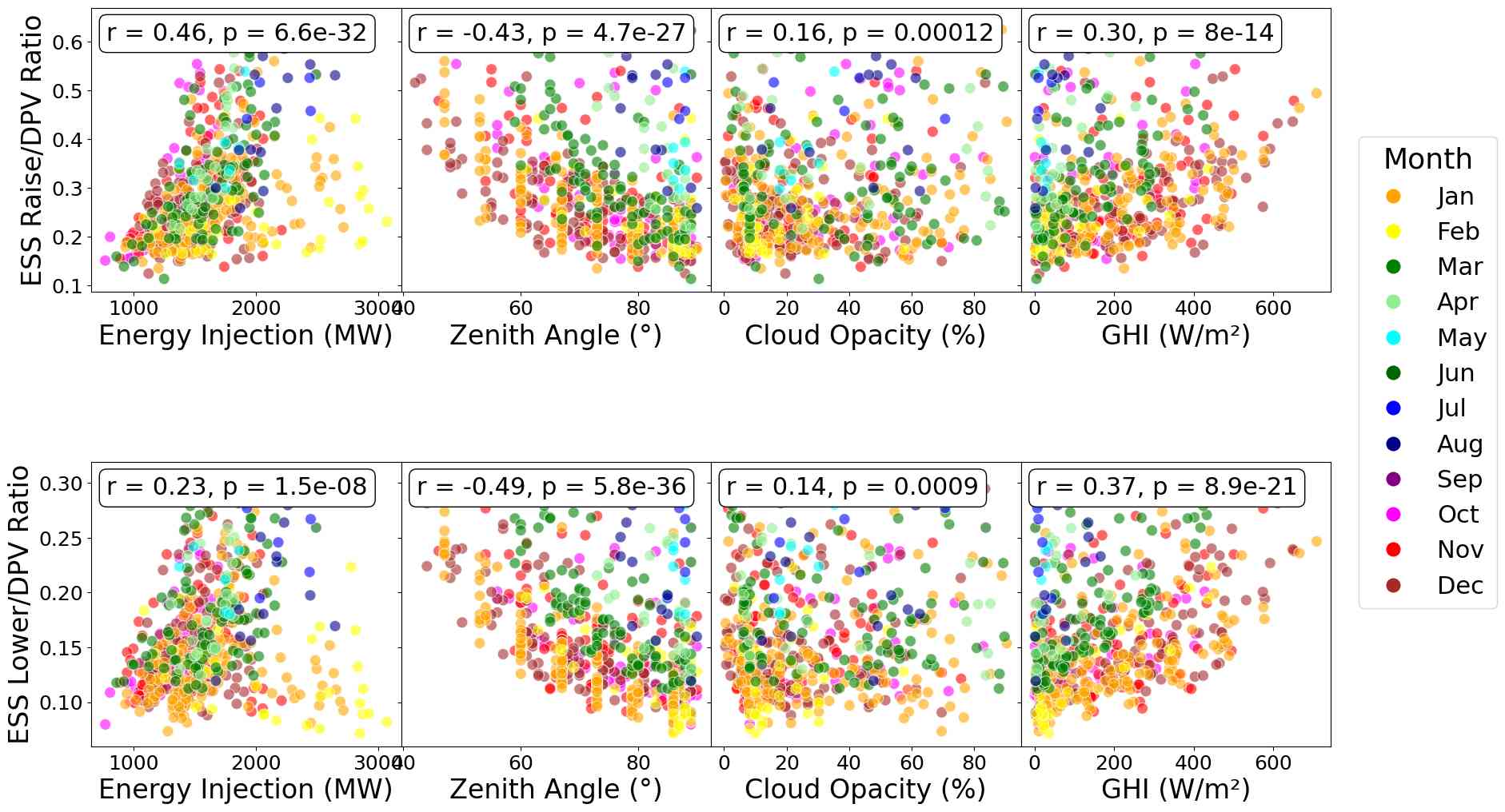}
    \caption{Correlation between weather variables and ESS Raise/DPV and ESS Lower/DPV ratios.}
    \label{fig:dpv_weather_impact}
\end{figure*}

Understanding how various market and environmental factors drive the ESS ratios is critical in predicting when the grid will need further support from ESS. According to Fig. \ref{fig:dpv_weather_impact}, the dominant Environmental and market variables that drive the main ESS activations are energy injection, zenith angle, cloud opacity, and Global Horizontal Irradiance (GHI).

The correlation coefficient, $r$, expresses the strength and direction of such relationships, and the p-value, denoted by $p$, yields the statistical significance; the smaller that value is, the greater the confidence in the result:

\newpage
\begin{enumerate}[leftmargin=*]
    \item {\em Energy Injection}: There exists a moderate positive relation between energy injection in MW and ESS ratios (r = 0.46 for Raise and r = 0.23 for Lower), which further provides evidence that higher injection, corresponding to high energy demand or energy generation surplus, provides more frequent ESS activation for raise as well as lower services.
    \item {\em Zenith Angle} It is to be noted that the zenith angle, a variable describing the position of the sun in the sky-is strongly negatively correlated with ESS ratios, its correlation coefficient for Raise is r = -0.43 and r = -0.49 for Lower. It is sensible to interpret this as the low sun-in other words, early morning and late afternoon-the grid relies more on ESS services.
    \item {\em Cloud Opacity:} Cloud opacity represents the extent of cloud cover ; cloud opacity is positively correlated with ESS demand ratios with a modest magnitude of r = 0.16 for Raise and r = 0.14 for Lower. This makes sense as with more cloud cover, less renewable generation by DPV exists hence creating an increased demand of ESS to make up for the shortfall in renewable generation.
    \item {\em GHI:} The correlation between the GHI vs ESS ratios is one of the strongest: r = 0.30 for Raise, and r = 0.37 for Lower. That is indicative of the fact that with increasing solar irradiance, higher generation of DPV occurs; thus, there is a reduction in the need for increased activation of ESS.
\end{enumerate}

These findings highlight important vulnerabilities in grid stability: {\em cloudy and low-sun conditions} increase the strain on ESS, while {\em clear skies and high GHI} contribute to a more stable grid with reduced ESS demand. Such insights point to the potential for timing disruptions, such as cyberattacks, to coincide with weather-related grid weaknesses for maximum impact.

\clearpage

\aptLtoX[graphic=no,type=html]{\begin{table*}[h!]
\tabcolsep3pt\centering 
{\scriptsize
\begin{tabular}{cc|ccccc|ccccc|cc|cc}
\hline
\multirow{3}{*}{Facility Number} & \multirow{3}{*}{Type} &
\multicolumn{5}{c|}{Max Capacity [MW] (or [MWs] for RoCoF Control)} &
\multicolumn{5}{c|}{Utilization Rate (Oct 23 - Sept 24)} 
& \multirow{3}{0.7cm}{\centering Speed Factor} & \multirow{3}{0.8cm}{\centering RoCoF Ride-Through} &
\multicolumn{2}{c}{Outages [hours]} \\ \cline{3-12}
& &
\ {Regulation} &
{Regulation} &
{Contingency} &
{Contingency} &
{RoCoF} &
\ {Regulation} &
{Regulation} & 
{Contingency} &
{Contingency} & 
{RoCoF} & & & \multicolumn{2}{c}{(Oct 23 - Sept 24)} \\ \cline{15-16}
 & &
Raise &
Lower &
Raise &
Lower &
Control &
Raise &
Lower &
Raise &
Lower &
Control\ & & & \ {Planned} & { Unplanned} \\ 
\hline
Facility 1 & Gas & 30  & 30 & & 58 & 1085 & 0.02 & 0.45 & 0.25 & & 0.77 & & 0.25 & 3320 & 2383 \\ 
\rowcolor[RGB]{255,182,193} Facility 2 & Gas & 30  & 30 & 6  & 65 & 1085 & 0.02 & 0.58 & 0.47 & 0.31 & 0.91 & 6 & 0.25 &  6780 & 3477 \\
Facility 3 & Distillate  & 30  & 30 & 55 & 60 & 1494 & 0.02 & 0.00 & 0.41 & 0.01 & 0.56 & 15 & 0.25 &  2122 & 376\\ 
Facility 4 & Distillate  & 30  & 30 & 60 & 60 & 1494 & 0.01 & 0.01 & 0.24 & 0.01 & 0.38 & 15 & 0.25 &  3091 & 2480\\ 
Facility 5 & Coal &  &  & 15 &  & 1077 &  &  & 0.20 & & 0.90 & 1 & 0.25 &4170 &  289\\
Facility 6 & Coal &  &  & 15 &  & 1077 &  &  & 0.14 & & 0.82 & 0.5 & 0.25 & 5704 & 132\\
Facility 7 & Gas & 40  & 40 & & & & 0.01 & 0.02 & & & & & 0.25 & 3847 & 303 \\
Facility 8 & Gas & 30  & 30 & 30  & 26.6 & 2686 & 0.03 & 0.46 & 0.02 & & 0.84 & 15 & 0.6 & 3946 & 62 \\
Facility 9 & Gas & 30  & 30 & & & 2176 & 0.02 & 0.13 & & & 0.31 & & 0.25 & 456 & 113\\
Facility 10 & Coal &  &  & 24.7  & 16.5 & 1196.3 & & & 0.03 & 0.03 & 0.36 & 10 & 0.25 & 4286 & 2448\\
Facility 11 & Distillate & 75  & 75 & 76  & 76 & 1092 & 0.03 & 0.01 & 0.07 & 0.01 & 0.12 & 10 & 0.25 & 113 & 456 \\
Facility 12 & Distillate & 75  & 75 & 76  & 76 & 1092 & 0.02 & 0.01 & 0.03 & 0.01 & 0.07 & 15 & 0.25 & 9989 & 92\\
\rowcolor[RGB]{255,182,193} Facility 13 & \ \ \ Battery \ \ \  & 100  & 100 & 50  & 50 & & 0.16 & 0.34 & 0.32 & 0.13 & & 0.5 & 0.25 & 2217 & 3585 \\
Facility 14 & Distillate & 80  & 80 & 52.5  & 52.5 & 222.6 & 0.19 & 0.24 & 0.46 & 0.05 & 0.83 & 10 & 0.25 & 4113 & 7910 \\
Facility 15 & Distillate & 80  & 80 & 52.5  & 52.5 & 222.6 & 0.23 & 0.12 & 0.54 & 0.04 & 0.85 & 10 & 0.25 & 4216 & 4981\\
Facility 16 & Coal &  &  & 21.2  & 21.2 & 880 & & & 0.29 & 0.35 & 0.66 & 6 & 0.25 & 1471 & 1514 \\
Facility 17 & Coal &  &  & 22.7  & 22.7 & 880 & & & 0.48 & 0.37 & 0.78 & 3 & 0.25 & 3706 & 986 \\
Facility 18 & Coal &  &  & 22.7  & 22.7 & 880 & & & 0.37 & 0.47 & 0.89 & 6 & 0.25 & 3946 & 10827\\
Facility 19 & Distillate & 28.5  & 28.5 & 12.9  & 12.9 & 248 & 0.02 & 0.01 & 0.19 & 0.01 & 0.21 & 3 & 0.25 & 6189 & 924\\
Facility 20 & Distillate & 28.5  & 28.5 & 12.9  & 12.9 & 248 & 0.06 & 0.01 & 0.14 & 0.00 & 0.16 & 3 & 0.25 & 3264 & 564\\
Facility 21 & Distillate & 29.3  & 29.3 & 15.7  & 15.7 & 276 & 0.08 & 0.01 & 0.18 & 0.01 & 0.21 & 3 & 0.25  & 5142 & 1791\\
Facility 22 & Distillate & 29.3  & 29.3 & 15.7  & 15.7 & 276 & 0.05 & 0.01 & 0.12 & 0.01 & 0.13 & 3 & 0.25 & 5644 & 1791\\
Facility 23 & Distillate & 29.3  & 29.3 & 11.4  & 11.4 & 276 & 0.11 & 0.01 & 0.22 & 0.01 & 0.25 & 3 & 0.25 & 5538 & 4001 \\
Facility 24 & Distillate & 29.3  & 29.3 & 11.4  & 11.4 & 276 & 0.11 & 0.01 & 0.24 & 0.01 & 0.26 & 3 & 0.25 & 2699 & 1370 \\
Facility 25 & Gas & 50  & 50 & 45.6  & 45.6 & 988 & 0.09 & 0.01 & 0.13 & 0.01 & 0.13 & 3 & 0.25 & 4638 & 7246  \\
\rowcolor[RGB]{255,182,193} Facility 26 & Gas & 50  & 50 & 37.7  & 37.7 & 988 & 0.46 & 0.02 & 0.64 & 0.02 & 0.72 & 3 & 0.25 & 1866 & 719 \\
Facility 27 & Gas & 50  & 50 & 55.9  & 55.9 & 1194 & 0.09 & 0.02 & 0.42 & 0.01 & 0.49 & 3 & 0.25 & 2199 & 2380\\
Facility 28 & Gas & 15  & 15 & & & & 0.01 & 0.01 & & & & & 0.25 & 570 & 387 \\
Facility 29 &  &  &   &  & 63 &  &  &  &  & 0.81 &  &  & 0.2 &  2359 & 4892 \\
\hline
\multicolumn{2}{c|}{Overall} & 969.2 & 969.2 & 861.5 & 878.9 & 23409.5 &  &  & & & & &  \\
\hline
\end{tabular}%
\\[10pt]
} 
\caption{Frequency Co-optimised
Essential System Services (ESS) facilities in an unnamed Australian state.} \label{tab:facilities}
\vspace{10pt}
\end{table*}
}{\begin{table*}[h!]
\centering 
{\scriptsize
\begin{tabular}{@{ }c@{\ }c@{|}c@{\ }c@{\ }c@{\ }c@{\ }c@{|}c@{\ }c@{\ }c@{\ }c@{\ }c@{|}c@{\ }c@{|}c@{}c@{}}
\hline
\multirow{3}{*}{Facility Number} & \multirow{3}{*}{Type} &
\multicolumn{5}{c|}{Max Capacity [MW] (or [MWs] for RoCoF Control)} &
\multicolumn{5}{c|}{Utilization Rate (Oct 23 - Sept 24)} 
& \multirow{3}{0.7cm}{\centering Speed Factor} & \multirow{3}{0.8cm}{\centering RoCoF Ride-Through} &
\multicolumn{2}{c}{Outages [hours]} \\ \cline{3-12}
& &
\ {Regulation} &
{Regulation} &
{Contingency} &
{Contingency} &
{RoCoF} &
\ {Regulation} &
{Regulation} & 
{Contingency} &
{Contingency} & 
{RoCoF} & & & \multicolumn{2}{c}{(Oct 23 - Sept 24)} \\ \cline{15-16}
 & &
Raise &
Lower &
Raise &
Lower &
Control &
Raise &
Lower &
Raise &
Lower &
Control\ & & & \ {Planned} & { Unplanned} \\ 
\hline
Facility 1 & Gas & 30  & 30 & & 58 & 1085 & 0.02 & 0.45 & 0.25 & & 0.77 & & 0.25 & 3320 & 2383 \\ 
\rowcolor[RGB]{255,182,193} Facility 2 & Gas & 30  & 30 & 6  & 65 & 1085 & 0.02 & 0.58 & 0.47 & 0.31 & 0.91 & 6 & 0.25 &  6780 & 3477 \\
Facility 3 & Distillate  & 30  & 30 & 55 & 60 & 1494 & 0.02 & 0.00 & 0.41 & 0.01 & 0.56 & 15 & 0.25 &  2122 & 376\\ 
Facility 4 & Distillate  & 30  & 30 & 60 & 60 & 1494 & 0.01 & 0.01 & 0.24 & 0.01 & 0.38 & 15 & 0.25 &  3091 & 2480\\ 
Facility 5 & Coal &  &  & 15 &  & 1077 &  &  & 0.20 & & 0.90 & 1 & 0.25 &4170 &  289\\
Facility 6 & Coal &  &  & 15 &  & 1077 &  &  & 0.14 & & 0.82 & 0.5 & 0.25 & 5704 & 132\\
Facility 7 & Gas & 40  & 40 & & & & 0.01 & 0.02 & & & & & 0.25 & 3847 & 303 \\
Facility 8 & Gas & 30  & 30 & 30  & 26.6 & 2686 & 0.03 & 0.46 & 0.02 & & 0.84 & 15 & 0.6 & 3946 & 62 \\
Facility 9 & Gas & 30  & 30 & & & 2176 & 0.02 & 0.13 & & & 0.31 & & 0.25 & 456 & 113\\
Facility 10 & Coal &  &  & 24.7  & 16.5 & 1196.3 & & & 0.03 & 0.03 & 0.36 & 10 & 0.25 & 4286 & 2448\\
Facility 11 & Distillate & 75  & 75 & 76  & 76 & 1092 & 0.03 & 0.01 & 0.07 & 0.01 & 0.12 & 10 & 0.25 & 113 & 456 \\
Facility 12 & Distillate & 75  & 75 & 76  & 76 & 1092 & 0.02 & 0.01 & 0.03 & 0.01 & 0.07 & 15 & 0.25 & 9989 & 92\\
\rowcolor[RGB]{255,182,193} Facility 13 & \ \ \ Battery \ \ \  & 100  & 100 & 50  & 50 & & 0.16 & 0.34 & 0.32 & 0.13 & & 0.5 & 0.25 & 2217 & 3585 \\
Facility 14 & Distillate & 80  & 80 & 52.5  & 52.5 & 222.6 & 0.19 & 0.24 & 0.46 & 0.05 & 0.83 & 10 & 0.25 & 4113 & 7910 \\
Facility 15 & Distillate & 80  & 80 & 52.5  & 52.5 & 222.6 & 0.23 & 0.12 & 0.54 & 0.04 & 0.85 & 10 & 0.25 & 4216 & 4981\\
Facility 16 & Coal &  &  & 21.2  & 21.2 & 880 & & & 0.29 & 0.35 & 0.66 & 6 & 0.25 & 1471 & 1514 \\
Facility 17 & Coal &  &  & 22.7  & 22.7 & 880 & & & 0.48 & 0.37 & 0.78 & 3 & 0.25 & 3706 & 986 \\
Facility 18 & Coal &  &  & 22.7  & 22.7 & 880 & & & 0.37 & 0.47 & 0.89 & 6 & 0.25 & 3946 & 10827\\
Facility 19 & Distillate & 28.5  & 28.5 & 12.9  & 12.9 & 248 & 0.02 & 0.01 & 0.19 & 0.01 & 0.21 & 3 & 0.25 & 6189 & 924\\
Facility 20 & Distillate & 28.5  & 28.5 & 12.9  & 12.9 & 248 & 0.06 & 0.01 & 0.14 & 0.00 & 0.16 & 3 & 0.25 & 3264 & 564\\
Facility 21 & Distillate & 29.3  & 29.3 & 15.7  & 15.7 & 276 & 0.08 & 0.01 & 0.18 & 0.01 & 0.21 & 3 & 0.25  & 5142 & 1791\\
Facility 22 & Distillate & 29.3  & 29.3 & 15.7  & 15.7 & 276 & 0.05 & 0.01 & 0.12 & 0.01 & 0.13 & 3 & 0.25 & 5644 & 1791\\
Facility 23 & Distillate & 29.3  & 29.3 & 11.4  & 11.4 & 276 & 0.11 & 0.01 & 0.22 & 0.01 & 0.25 & 3 & 0.25 & 5538 & 4001 \\
Facility 24 & Distillate & 29.3  & 29.3 & 11.4  & 11.4 & 276 & 0.11 & 0.01 & 0.24 & 0.01 & 0.26 & 3 & 0.25 & 2699 & 1370 \\
Facility 25 & Gas & 50  & 50 & 45.6  & 45.6 & 988 & 0.09 & 0.01 & 0.13 & 0.01 & 0.13 & 3 & 0.25 & 4638 & 7246  \\
\rowcolor[RGB]{255,182,193} Facility 26 & Gas & 50  & 50 & 37.7  & 37.7 & 988 & 0.46 & 0.02 & 0.64 & 0.02 & 0.72 & 3 & 0.25 & 1866 & 719 \\
Facility 27 & Gas & 50  & 50 & 55.9  & 55.9 & 1194 & 0.09 & 0.02 & 0.42 & 0.01 & 0.49 & 3 & 0.25 & 2199 & 2380\\
Facility 28 & Gas & 15  & 15 & & & & 0.01 & 0.01 & & & & & 0.25 & 570 & 387 \\
Facility 29 &  &  &   &  & 63 &  &  &  &  & 0.81 &  &  & 0.2 &  2359 & 4892 \\
\hline
\multicolumn{2}{c|}{Overall} & 969.2 & 969.2 & 861.5 & 878.9 & 23409.5 &  &  & & & & &  \\
\hline
\end{tabular}%
\\[10pt]
} 
\caption{Frequency Co-optimised
Essential System Services (ESS) facilities in an unnamed Australian state.} \label{tab:facilities}
\vspace{10pt}
\end{table*}}

\newpage

\aptLtoX[graphic=no,type=html]{\begin{figure*}[t!]
\centering
\includegraphics[scale=0.4]{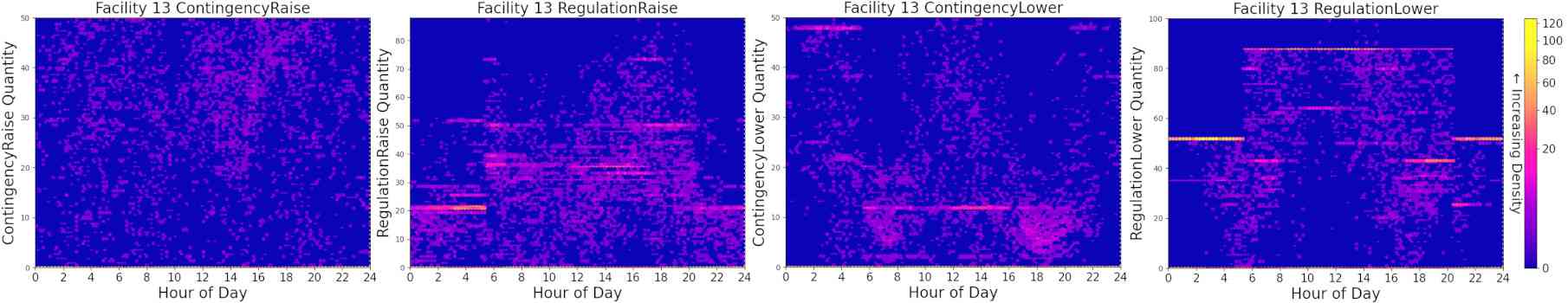}  \vspace{-10pt}
\caption{The dispatch statistics of Facility 13 for  Regulation and Contingency services in Oct 2023- Sept 2024.} 
\label{fig:kwinana} %\vspace{-15pt}
\end{figure*}
\begin{figure*}
\includegraphics[scale=0.4]{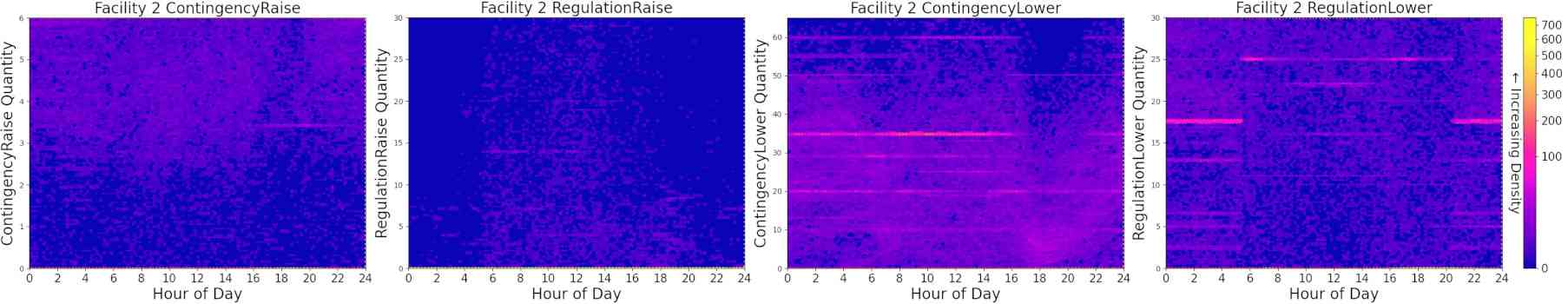}  \vspace{-10pt}
\caption{The dispatch statistics of Facility 2 for  Regulation and Contingency services in Oct 2023- Sept 2024.}
\label{fig:alinta} %\vspace{-15pt}
\end{figure*}
\begin{figure*}
\includegraphics[scale=0.2]{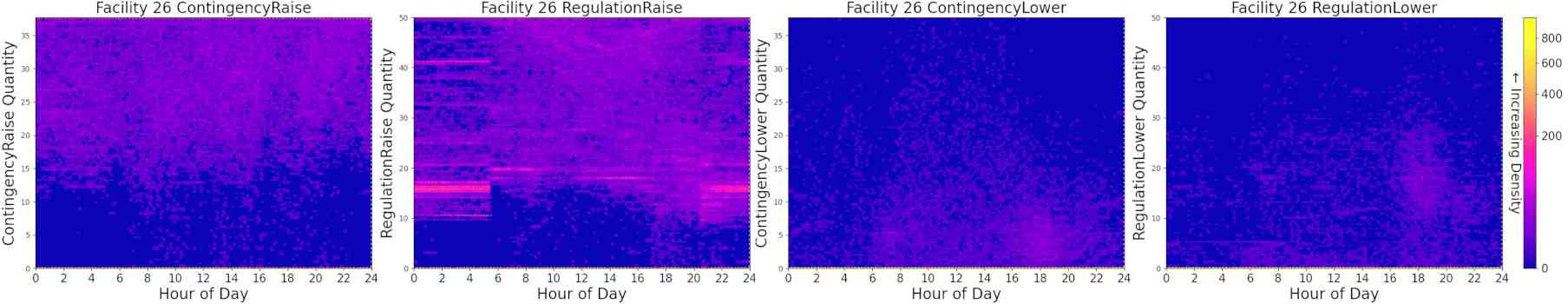}  \vspace{-10pt}
\caption{The dispatch statistics of Facility 26 for  Regulation and Contingency services in Oct 2023- Sept 2024.}
\label{fig:pinjar} %\vspace{-15pt}
%\vspace{20pt}
\end{figure*}}{
\begin{figure*}[t!]
\centering
\includegraphics[scale=0.25]{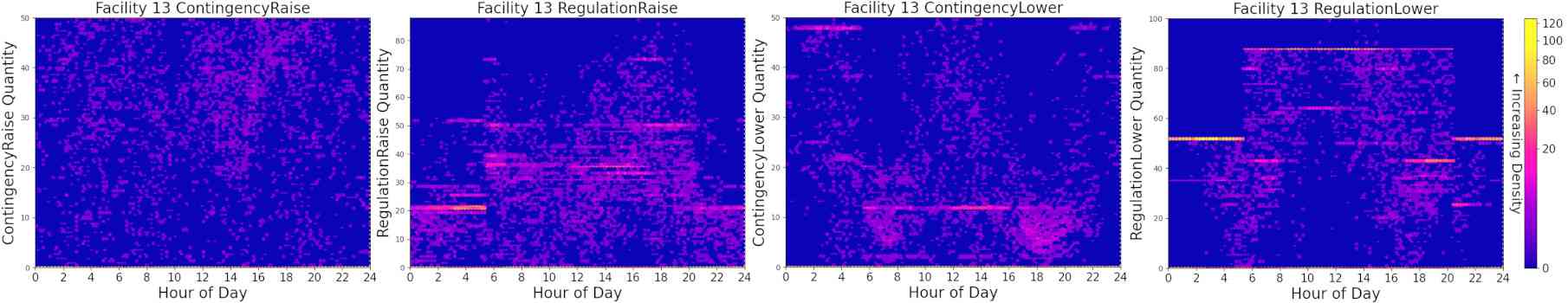}  \vspace{-10pt}
\caption{The dispatch statistics of Facility 13 for  Regulation and Contingency services in Oct 2023- Sept 2024.} 
\label{fig:kwinana} %\vspace{-15pt}
%\bigskip
\includegraphics[scale=0.25]{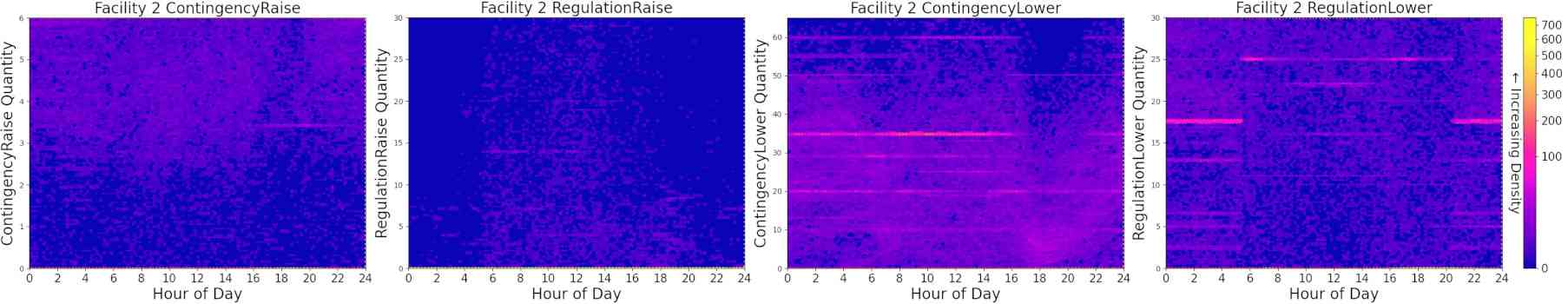}  \vspace{-10pt}
\caption{The dispatch statistics of Facility 2 for  Regulation and Contingency services in Oct 2023- Sept 2024.}
\label{fig:alinta} %\vspace{-15pt}
%\bigskip
\includegraphics[scale=0.25]{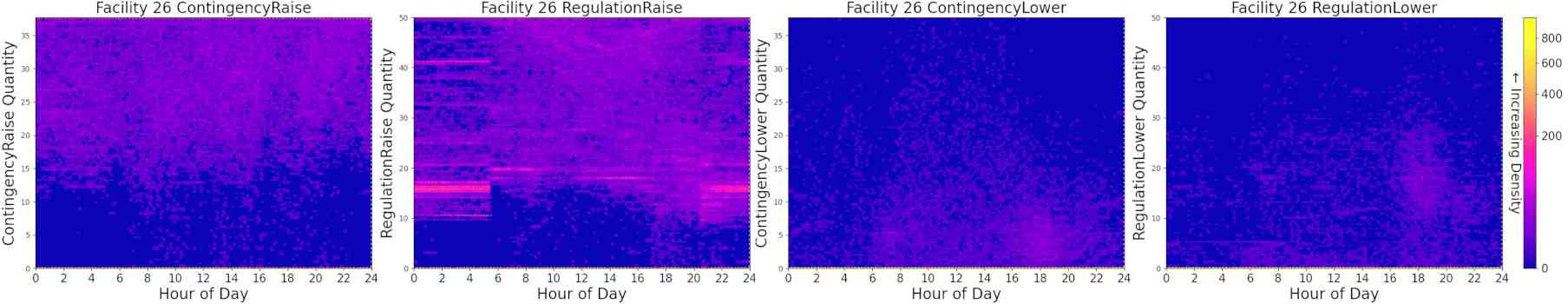}  \vspace{-10pt}
\caption{The dispatch statistics of Facility 26 for  Regulation and Contingency services in Oct 2023- Sept 2024.}
\label{fig:pinjar} %\vspace{-15pt}
%\vspace{20pt}
\end{figure*}}

\newpage
% \clearpage

\end{document}